\documentclass[11pt]{article}
\usepackage{amsmath}
\usepackage{graphicx}
\usepackage[usenames,dvipsnames,table]{xcolor}
\usepackage{amssymb}
\usepackage{float}
\usepackage{scrtime}
\usepackage{fancyhdr}
\usepackage{subfigure}
\usepackage{authblk}
\usepackage{lipsum}
\usepackage{rotating}
\usepackage{floatpag}
\usepackage{varioref}
\usepackage{slashed}
\usepackage{enumerate}
\usepackage{listings}
\usepackage{cancel}
\usepackage{wrapfig}

\usepackage{caption}
\DeclareCaptionFont{white}{\color{white}}
\DeclareCaptionFormat{listing}{%
  \parbox{\textwidth}{\colorbox{gray}{\parbox{\textwidth}{#1#2#3}}\vskip-4pt}}
\usepackage[colorlinks=true
,urlcolor=blue
,anchorcolor=blue
,citecolor=blue
,filecolor=blue
,linkcolor=black
,menucolor=blue
,pagecolor=blue
,linktocpage=true
]{hyperref}

\usepackage[numbers,sort&compress]{natbib}

\setlipsumdefault{131}

\newcommand{\maton}{\texttt{maton}}

\newcommand{\code}[1]{{\small{\texttt{#1}}}}
\newcommand{\filename}[1]{{\small{\texttt{#1}}}}
\newcommand{\ra}{\ensuremath{\rightarrow}}

\newcommand{\SO}[1]{\ensuremath{\mathrm{SO}(#1)}}

\newcommand{\bs}[1]{\ensuremath{\boldsymbol{#1}}}

\labelformat{section}{Section #1}
\labelformat{subsection}{Section #1}
\labelformat{equation}{Eq.~(#1)}
\labelformat{figure}{Fig.~#1}
\labelformat{subfigure}{Fig.~\thefigure#1}
\labelformat{table}{Tab.~#1}
\labelformat{lstlisting}{Listing~#1}

\def\a{\alpha}

\def\be{\begin{equation}}
\def\ee{\end{equation}}
\def\bea{\begin{eqnarray}}
\def\eea{\end{eqnarray}}

\graphicspath{{./figures/}}

\begin{document}

\floatpagestyle{plain}

\pagenumbering{roman}

\renewcommand{\headrulewidth}{0pt}
% \lhead{
% \today \quad \thistime
% }
\rhead{
OHSTPY-HEP-T-13-002\\
LPSC-13183\\
CETUP2013-006\\
}
\fancyfoot{}

\title{\huge \bf{LHC Phenomenology of\\ \bs{\SO{10}} Models with Yukawa Unification}}

\author[$\dag$]{Archana Anandakrishnan}
\author[$\dag$]{B. Charles Bryant}
\author[$\dag$]{Stuart Raby}
\author[$^\ddag$]{Ak\i{}n Wingerter}

\affil[$\dag$]{\em Department of Physics, The Ohio State University,\newline
191 W.~Woodruff Ave, Columbus, OH 43210, USA \enspace\enspace\enspace\enspace \medskip}

\affil[$^\ddag$]{\em Laboratoire de Physique Subatomique et de Cosmologie,\newline
UJF Grenoble 1, CNRS/IN2P3, INPG,\newline
53 Avenue des Martyrs, F-38026 Grenoble, France\enspace\enspace\enspace\enspace\enspace}	

\maketitle
\thispagestyle{fancy}

\begin{abstract}\normalsize\parindent 0pt\parskip 0pt
In this paper we study an SO(10) SUSY GUT with Yukawa unification for the third generation.  We perform a global $\chi^2$ analysis given to obtain the GUT boundary conditions consistent with 11 low energy observables, including the top, bottom and tau masses.  We assume a universal mass, $m_{16}$, for squarks and  sleptons and a universal gaugino mass, $M_{1/2}$. We then analyze the phenomenological consequences for the LHC for 15 benchmark models with fixed $m_{16} = 20$ TeV and with varying values of the gluino mass.   The goal of the present work is to (i) evaluate the lower bound on the gluino mass in our model coming from the most recent published data of CMS and (ii) to compare this bound with similar bounds obtained by CMS using simplified models.  The bottom line is that the bounds coming from the same sign di-lepton analysis are comparable for our model and the simplified model studied assuming $\mathcal{B}(\tilde g \rightarrow t \ \bar t \ \tilde \chi^0_1) = 100\%$.   However the bounds coming from the purely hadronic analyses for our model are 10 - 20\% lower than obtained for the simplified models.  This is due to the fact that for our models the branching ratio for the decay $\tilde g \rightarrow g \ \tilde \chi^0_{1,2}$ is significant.  Thus there are significantly fewer b-jets.   We find a lower bound on the gluino mass in our models with $M_{\tilde g} \gtrsim 1000$ GeV.   Finally,  there is a theoretical upper bound on the gluino mass which increases with the value of $m_{16}$.   For $m_{16} \leq 30$ TeV, the gluino mass satisfies $M_{\tilde g} \leq 2.8$ TeV at 90\% CL.  Thus, unless we further increase the amount of fine-tuning, we expect gluinos to be discovered at LHC 14.
\end{abstract}

\clearpage
\newpage

\pagenumbering{arabic}
%\setcounter{page}{1}

%%%%%%%%%%%%%%%%%%%%%%%%%%%%% INTRODUCTION %%%%%%%%%%%%%%%%%%%%%%%%%%%%%%%
\section{Introduction}

In a previous paper, \cite{Anandakrishnan:2012tj}, some of the present authors
performed a global $\chi^2$ analysis on the Dermisek-Raby [DR] model
\cite{Dermisek:2005sw,Albrecht:2007ii}.  The DR model has $SO(10)$ Yukawa
unification for the third family of quarks and leptons and a hierarchical
set of Yukawa matrices for the two lighter families which is determined by a
hierarchical breaking of a $D_3$ family symmetry.  This model has 24 arbitrary
parameters (22 gauge, Yukawa and soft SUSY breaking parameters defined at the
GUT scale with two additional parameters, $\mu, \tan\beta$ fixed at the weak
scale), see \ref{tab:parameters}.     With a $\chi^2$ function including 36 low
energy observables we found the best fit with $\chi^2/d.o.f. = 2$ for a value of
the universal squark and slepton mass parameter, $m_{16} = 20$ TeV and a lower
bound on the gluino mass, $M_{\tilde g} = 850$ GeV,  imposed via a $\chi^2$
penalty.  In the previous paper it was also demonstrated that the SUSY particle
spectrum was constrained solely by a third family analysis.  It is this analysis
we use in the present paper.

Note, the gluino mass prefers to be even smaller than 850 GeV, but a lower bound
was imposed to be roughly consistent with recent LHC SUSY search bounds.  The
goal of this paper is to find the actual limits on the SUSY spectrum consistent
with Yukawa unification by directly comparing with the most recent LHC SUSY
searches.   The SUSY bounds which are most relevant to our SUSY spectrum are
those obtained using simplified models.   In particular,  we find the squarks
and sleptons of the first two families to have a large mass of order $m_{16}$
which we take to be between 10 and 30 TeV.   The third family of squarks and
sleptons, however, are significantly lighter and finally the gauginos are even
lighter.   Therefore, the simplified models which might come closest to
constraining our results are those in which gluinos are pair produced at the LHC
and then decay with 100\% branching ratio  $\tilde g \rightarrow t \ \bar t \
\tilde \chi^0_1$ or $\tilde g \rightarrow b \ \bar b \ \tilde \chi^0_1$.  The
lower bound on the gluino mass from CMS
\cite{Chatrchyan:2013wxa,Chatrchyan:2012paa,Chatrchyan:2013lya} and ATLAS
\cite{ATLAS-CONF-2013-007,ATLAS-CONF-2013-054,ATLAS-CONF-2013-047} searches are
of order $M_{\tilde g} \geq 1 - 1.2$ TeV.    In this paper we will not impose a
lower bound on the gluino mass by hand, as we had done in our previous analysis.   In \ref{fig:CL-intervals} we plot
$\chi^2$ as a function of the gluino mass in a third family analysis.  Since
$M_{1/2}$ and $m_{16}$ are fixed at the GUT scale,  there are 9 arbitrary parameters and 11
low energy observables or 2 degrees of freedom.  The gluino mass wants to be low
with the best fit at $M_{\tilde g} \simeq 600$ GeV.  We also see that $\chi^2$
increases as the gluino mass increases.  Thus as discussed in
\cite{Anandakrishnan:2012tj} there is an upper bound on the gluino mass such
that at 90\% CL we have $M_{\tilde g} \lesssim 2.0$ TeV.\footnote{This upper bound depends on the fact that
$m_{16} = 20$ TeV.  For larger $m_{16}$ the upper bound on the gluino mass increases (See \ref{app:gluino} for more details).}
In fact, we shall show that the 68\% C.L. interval is not totally excluded by
LHC searches.   We find that gluinos are allowed with mass approximately 20\%
smaller than presently excluded using some simplified model analyses.  Nevertheless we find the strongest constraint comes
from the same sign di-lepton analysis and the hadronic $\Delta \phi$ analysis which both give $M_{\tilde g} \gtrsim 1$ TeV.

The paper is organized as follows.  In \ref{sec:benchmark} we describe the model
and the benchmark points used in the present analysis.
Our analysis uses several of the SUSY tools in the literature, such as \code{Prospino},
\code{Pythia}, \code{Delphes}, \code{SDecay} and \code{ROOT}, in addition to the code \code{Maton}, developed
by Dermisek and revised by Anandakrishnan and Wingerter to be applicable to the
hierarchy of the low energy SUSY scales as discussed in \ref{sec:procedure}.
We calculate the constraints on the gluino mass using the  same sign di-lepton
and hadronic analyses of CMS.  These are the most relevant ones for
our model, since there is sufficient information to abstract the 95\% C.L. upper
limit on the number of SUSY events allowed in each signal region.  The analysis
using the CMS data for same sign di-leptons is found in
\ref{sec:dilepton} and jets plus missing $E_T$ is in \ref{sec:hadronic}.
Finally a summary of our results and conclusions is given in
\ref{sec:conclusions}.

%\clearpage
\section{Benchmark Models}
\label{sec:benchmark}

In Ref. \cite{Anandakrishnan:2012tj}, it was found that the soft SUSY spectrum was
completely determined by fitting 11 low energy observables, $M_W$, $M_Z$, $G_F$,
$\alpha_{em}^{-1}$, $\alpha_s(M_Z)$, $M_t$, $m_b(m_b)$, $M_{\tau}$,
$\mathcal{B}(B\rightarrow X_s \gamma)$, $\mathcal{B}(B_s \rightarrow \mu^+ \mu^-)$, and the
lightest Higgs mass, $M_h$. Therefore in this paper we use the third family
analysis exclusively with the 11 arbitrary parameters listed in
\ref{tab:parameters}.   The universal Yukawa coupling $\lambda$ appears in the
superpotential term ${\cal W} \supset \lambda \ 16_3 \ 10 \ 16_3$.

In the present analysis we neglect all Yukawa terms except for the third family
coupling, $\lambda$, and integrate out the right-handed neutrinos at the GUT
scale.   One might wonder whether neglecting the off-diagonal elements of the
Yukawa matrices, as considered in our three family analysis, and also not
allowing the right-handed neutrinos to be integrated out at lower scale, more
appropriate for the See-Saw mechanism, might effect significant deviations from
the earlier paper Ref. \cite{Anandakrishnan:2012tj}.  Nevertheless, we find that
within a few percent the results are the same.  In the former case, however, the
magnitude of the Higgs splitting parameter is all that needs adjusting, since
approximately 50\% of the Higgs splitting necessary for electroweak symmetry
breaking is generated as a threshold effect due to the right-handed neutrinos.

\begin{table}[h!]
\begin{center}
\renewcommand{\arraystretch}{1.2}
\scalebox{0.83}{
\begin{tabular}{|l||c|c||c|c||}
\hline
Sector &  Third Family Analysis & \# & Full three family Analysis & \# \\
\hline
gauge             & $\alpha_G$, $M_G$, $\epsilon_3$                   & 3  &
$\alpha_G$, $M_G$, $\epsilon_3$
& 3 \\
SUSY (GUT scale)  & $m_{16}$, $M_{1/2}$, $A_0$, $m_{H_u}$, $m_{H_d}$  & 5  &
$m_{16}$, $M_{1/2}$, $A_0$, $m_{H_u}$, $m_{H_d}$
& 5 \\
textures          & $\lambda$                                         & 1  &
$\epsilon$, $\epsilon'$, $\lambda$, $\rho$, $\sigma$, $\tilde \epsilon$, $\xi$
& 11\\
neutrino          &                                                   & 0  &
$M_{R_1}$, $M_{R_2}$, $M_{R_3}$
& 3 \\
SUSY (EW scale)   & $\tan \beta$, $\mu$                               & 2  &
$\tan \beta$, $\mu$
& 2 \\
\hline
Total \#          &                                                   & 11 &
                                                                            &
24\\
\hline
\end{tabular}
}
\caption{\footnotesize The model is defined by three gauge parameters,
$\alpha_{G}, M_{G}, \epsilon_3$; one large Yukawa coupling, $\lambda$; 5 SUSY
parameters defined at the GUT scale, $m_{16}$ (universal scalar mass for squarks
and sleptons), $M_{1/2}$ (universal gaugino mass), $m_{H_u}, \ m_{H_d}$ (up and
down Higgs masses), and $A_0$ (universal trilinear scalar coupling); $\mu, \
\tan\beta$ obtained at the weak scale by consistent electroweak symmetry
breaking. The full three family model has additional off-diagonal Yukawa
couplings, and includes 3 right-handed neutrino masses.}
\label{tab:parameters}
\end{center}
\end{table}

\begin{table}
\centering
\renewcommand{\arraystretch}{1.2}
\begin{tabular}{|l|l|l|l|l|l|l|}
\hline
          & YUa & YUb & YUc  & YUd &  YUe & YUf  \\
\hline
$M_{1/2}$ & 150 & 200 & 250  & 300 &  400 & 600  \\
%$ $  &  10 TeV  &  15 TeV &  20 TeV & 25 TeV & 30 TeV \\
%$A_0$  & -20.2 TeV  & -30.6 TeV & -41.1 TeV & -51.3 TeV & -61.6 TeV   \\
$\mu  $  & 869 & 889 & 824 & 878 & 975 & 664    \\
\hline
$\chi^2$ & 0.22 & 0.43 & 0.77 & 1.17 &  2.24 & 3.97 \\
\hline
$M_A$ & 2299 & 2323 & 2262 & 2300 &  2750 & 2120 \\
$m_{\tilde t_1}$ & 3775 & 3789 & 3695 & 3728 & 3775  & 3530 \\
$m_{\tilde b_1}$  & 4633 & 4654 & 4579 & 4607 & 4628 & 4526  \\
$m_{\tilde \tau_1}$  & 7865 & 7885 & 7834 & 7806 &  7890  & 7897 \\
$m_{\tilde\chi^0_1}$   & 129 & 151 & 172 & 194 &  238  & 325   \\
$m_{\tilde\chi^+_1}$  & 263 & 303 & 342 & 382 &  461  & 587   \\
$M_{\tilde g}$   & 801 & 932 & 1061 & 1186 &  1430  & 1890  \\
\hline
\end{tabular}
\caption{\footnotesize SUSY spectrum for six benchmark models corresponding to
the third-family analysis in Ref.~\cite{Anandakrishnan:2012tj} with the
qualifications detailed in the main text. All masses are in GeV. For each model,
we list the 11 defining parameters in \ref{app:benchmark}. The scale where the
first and second generation scalars are integrated out is fixed at $m_{16}=20$
TeV, and the universal trilinear coupling comes out close to $A_0=-41$ TeV in
each case as a result of the minimization. In the first column, we have assigned
names to the benchmark models that we will use in the following discussion to
refer to them. The third row gives the total $\chi^2$ and the rest of the lines are self-explanatory.}
\label{tab:benchmark-spectrum}
\end{table}

\begin{table}
\centering
\renewcommand{\arraystretch}{1.5}
\begin{tabular}{|l|r|r|r|r|r|r|}
\hline
YUa & $tb\widetilde{\chi}^\pm_1$ & $t\bar{t}\widetilde{\chi}^0_2$ & $g \widetilde{\chi}^0_2$ & &$ t\bar{t}\widetilde{\chi}^0_1$ & $b\bar{b}\widetilde{\chi}^0_1$ \\
& 52\% & 22\% & 19\% & &4.5\% & 0.07\%\\
\hline
YUb & $tb\widetilde{\chi}^\pm_1$ & $t\bar{t}\widetilde{\chi}^0_2$ & $ g \widetilde{\chi}^0_2$ &$ g \widetilde{\chi}^0_{(1,3,4)}$ &$ t\bar{t}\widetilde{\chi}^0_1$ & $ b\bar{b}\widetilde{\chi}^0_1$ \\
& 50\% & 27\% & 11\% &5.5\% & 4.5\% & 0.04\%\\
\hline
YUc & $ g \widetilde{\chi}^0_4$ & $ g \widetilde{\chi}^0_3$ & $ tb\widetilde{\chi}^\pm_1$ &$ g \widetilde{\chi}^0_2$  &$ t\bar{t}\widetilde{\chi}^0_1$ & $ b\bar{b}\widetilde{\chi}^0_1$ \\
& 38\% & 35\% & 14\% &8\% & 1.2\% & 0.006\%\\
\hline
YUd & $ g \widetilde{\chi}^0_4$ & $ g \widetilde{\chi}^0_3$ & $ tb\widetilde{\chi}^\pm_{(1,2)}$ & $ t\bar{t}\widetilde{\chi}^0_2$ &$ t\bar{t}\widetilde{\chi}^0_1$ & $ b\bar{b}\widetilde{\chi}^0_1$ \\
& 35\% & 33\% & 21\% &7\% & 1\% & 0.004\%\\
\hline
YUe & $ tb\widetilde{\chi}^\pm_2$ & $ g \widetilde{\chi}^0_{(3,4)}$ & $ t \bar{t}\widetilde{\chi}^0_{(2,3,4)}$ & $ tb\widetilde{\chi}^\pm_1$ &$ t\bar{t}\widetilde{\chi}^0_1$ & $ b\bar{b}\widetilde{\chi}^0_1$ \\
& 34\% & 34\% & 23\% &6\% & 0.5\% & 0.0002\%\\
\hline
YUf & $ t \bar{t}\widetilde{\chi}^0_4$ & $ tb\widetilde{\chi}^\pm_2 $ & $ t \bar{t}\widetilde{\chi}^0_{3}$ & $ tb\widetilde{\chi}^\pm_1$ &$ t\bar{t}\widetilde{\chi}^0_1$ & $ b\bar{b}\widetilde{\chi}^0_1 $ \\
& 31\% & 23\% & 19\% &14\% & 0.25\% & 0.000001\%\\
\hline
\end{tabular}
\caption{\footnotesize Gluino decay branching ratios into different final states for the six benchmark models given in \ref{tab:benchmark-spectrum}. For each model, we give the dominant branching fractions. The last two columns give the branching fractions of the gluino into $t \ \bar{t} \ \widetilde{\chi}^0_1$ and $b \ \bar{b} \ \widetilde{\chi}^0_1$ which correspond to the simplified scenarios studied by the CMS analyses. These rates were calculated using \code{SDECAY}.}
\label{tab:benchmark-decays}
\end{table}

We have to emphasize, however, one important difference with respect to the
previous analysis. In Ref.~\cite{Anandakrishnan:2012tj}, we incorporated
collider constraints by requiring that the gluino be greater than a given lower
bound, i.e.~while minimizing $\chi^2$, we introduced a penalty whenever the
gluino mass was less than this lower bound. These lower bounds were taken from
LHC analyses based on so-called simplified models. Although as a first guess,
bounds derived using simplified models may be useful, it turns out that in
practice the assumptions entering the analyses are rarely met.

%%%%%%%%%%%%%%% PLOT 1 %%%%%%%%%%%%%%%
\begin{figure}[h!]
\thisfloatpagestyle{empty}
\centering
\includegraphics[width=0.8\textwidth, trim=0ex 0ex 0ex 4ex,
clip]{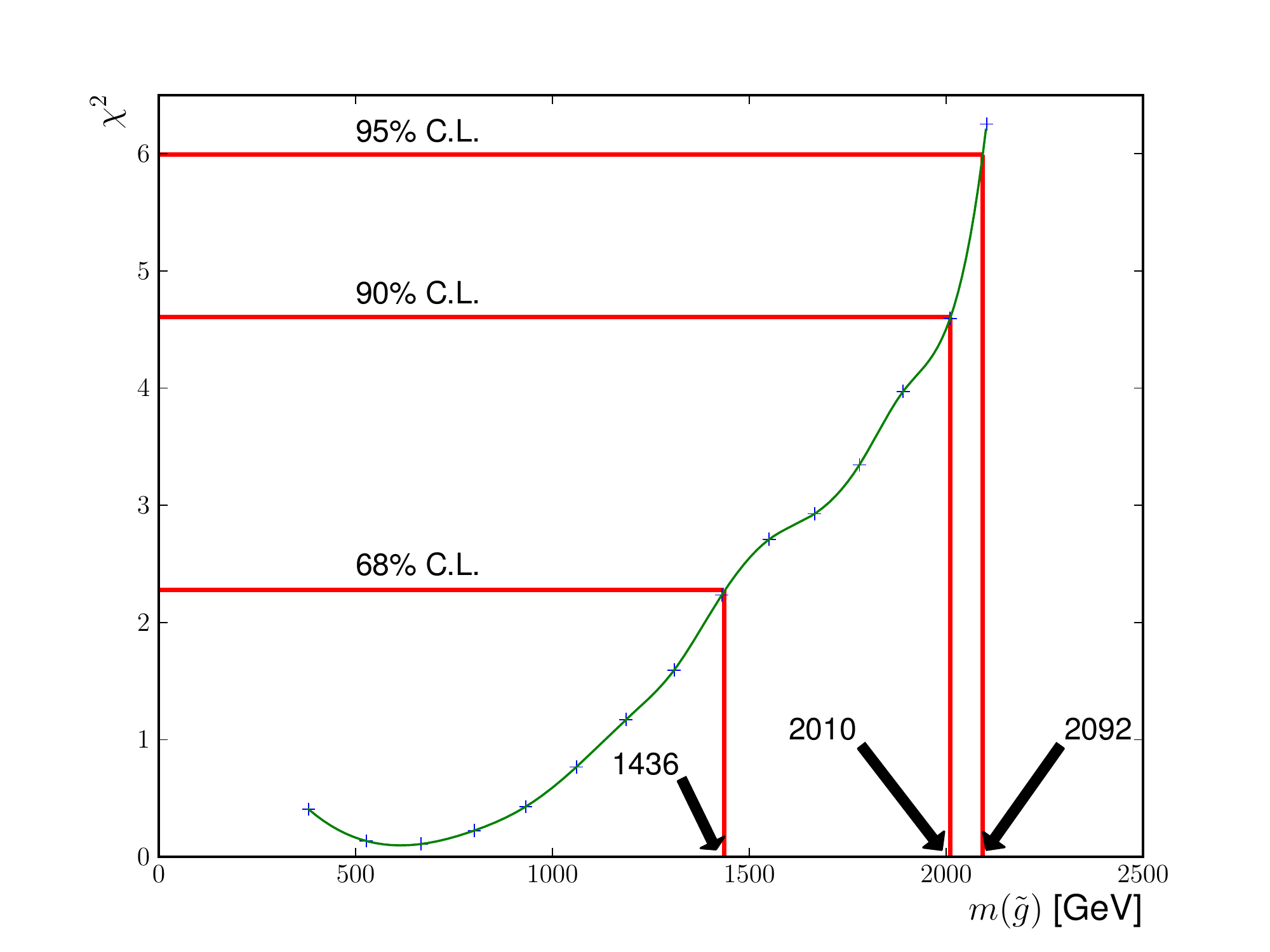}
\caption{\footnotesize The $\chi^2$ function for fixed $m_{16} = 20$ TeV and
different values of $M_{1/2}$ corresponding to the gluino masses indicated on
the $x$-axis. In minimizing $\chi^2$ for each point, we followed the same
procedure as in Ref.~\cite{Anandakrishnan:2012tj} except that we now do not
penalize for gluino masses smaller than a certain lower bound. The confidence
level intervals correspond to the $\chi^2$ distribution with two degrees of
freedom.}
\label{fig:CL-intervals}
\end{figure}
%%%%%%%%%%%%%%%%%%%%%%%%%%%%%%%%%%%%%%

From what we have discussed so far it follows that it does not make much sense
to constrain the $\chi^2$ minimization by requiring a lower bound on the gluino
mass derived from simplified models. As a consequence, we drop this requirement
and present in \ref{fig:CL-intervals} the corresponding $\chi^2$ as a function
of the gluino mass. More specifically, of the 11 parameters defining the
third-family SO(10) Yukawa unified model (see Tab.~1), we fix
$m_{16}=20$ TeV, and then minimize $\chi^2$ for 15 values of $M_{1/2}$ between
0 and 600 GeV by varying all the other parameters.  The $x$ axis in \ref{fig:CL-intervals} is
$M_{\tilde g}$.  The data points are
indicated in \ref{fig:CL-intervals} by crosses, and we have connected them using
a cubic spline interpolation to guide the eye. We calculate the 68\%, 90\% and
95\% confidence level intervals using the $\chi^2$ distribution for 2 degrees of
freedom; the corresponding values are indicated in \ref{fig:CL-intervals} by the
horizontal red lines. We find that they correspond to gluino masses of 1436,
2010, and 2092 GeV and are currently far from being excluded even by the
simplified model analyses. As benchmark models in \ref{tab:benchmark-spectrum}
we have chosen six models along this $\chi^2$ curve corresponding to a range of
different gluino masses.

In our models,
the first and second generation squarks are heavy, and the particles in the
spectrum that are accessible at the LHC are the the gluinos, the lightest
neutralino and the charginos (cf.~\ref{tab:benchmark-spectrum}). The simplified
models that come closest to our setup assume branching ratios of
$\mathcal{B}(\widetilde{g}\rightarrow t \ \bar{t} \ \widetilde{\chi}^0_1)=100\%$ or
$\mathcal{B}(\widetilde{g}\rightarrow b \ \bar{b} \ \widetilde{\chi}^0_1)=100\%$. For the benchmark models (\ref{tab:benchmark-spectrum}) the decay
branching ratios for the gluino are given in \ref{tab:benchmark-decays}. These clearly do not match the decay
branching ratios for any simplified model.

Fortunately, the ATLAS and CMS collaborations provide in many cases the raw data
necessary to reinterpret the searches for new physics in the context of more
elaborate models. E.g.~the model A1 in Ref.~\cite{Chatrchyan:2012paa} assumes
$\mathcal{B}(\widetilde{g}\rightarrow t\bar{t}\widetilde{\chi}^0_1)=100\%$, but
the expected number of SM events in each signal region (cmp.~Tab.~2 in the same
publication) allows us to derive exclusion bounds for our model on the basis of
not having observed a signal. This is in essence what we are doing in this
paper.   We will now move on to explain the procedure how to
derive bounds for these models from the recent CMS analyses.

%%%%%%%%%%%%%%%%%%%%%%%%%%%%% PROCEDURE %%%%%%%%%%%%%%%%%%%%%%%%%%%%%%%

%\clearpage
\section{Procedure}
\label{sec:procedure}

For clarity, we have summarized the analysis chain in \vref{fig:flowchart}. As
mentioned before, each benchmark model is defined in terms of 11 parameters
which we list for reference in \ref{tab:parameters}. We use the spectrum
generator \maton{} originally written by R.~Derm\'{i}\v{s}ek to find the
low-energy spectrum which is then written to a file in the SUSY Les Houches
Accord (SLHA) format.\footnote{ Note that storing the spectrum in a form that is
accessible by other programs is absolutely crucial for our analysis, since the
different components in the work chain need to communicate with each other in
order to produce data that can be directly compared to experiment.}  We emphasize that \maton{} has been modified to take into account
the decoupling of the first two families of squarks and sleptons.  In addition, we are greatful
to P. Slavich who has provided us with the code to calculate the light Higgs mass.
Finally, we have compared the output SUSY spectrum of \maton{} with SoftSUSY and the
results agree to within several percent.

\begin{table}
\begin{center}
\begin{tabular}{|c|c|c|c|}
\hline
Analysis & Luminosity & Signal Region & Reference \\
\hline
SS di-lepton & 10.5 & $ N_\mathrm{jet} \geq 4,\  N_{\mathrm{b-jet}} \geq 2, $  &
\cite{Chatrchyan:2012paa}\\
& & $ E_T^{\mathrm{miss}} > 120, \ H_T > 200$ & \\
\hline
$\alpha_T$ analysis (for Simplified models) & 11.7 & $ N_\mathrm{jet} \geq 4,\
N_{\mathrm{b-jet}} = 3,\ H_T > 875$  & \cite{Chatrchyan:2013lya}\\
   (for the benchmark models) & & \ $ N_\mathrm{jet} \geq 4,\
N_{\mathrm{b-jet}} = 2,\ 775 < H_T < 875$ &  \\
 \hline
$\Delta \phi$ analysis & 19.4 & $N_{\mathrm{b-jet}} \geq 3 ,\
E_T^{\mathrm{miss}} > 350, \ H_T > 1000$& \cite{Chatrchyan:2013wxa}\\
\hline
\end{tabular}
\caption{\label{tab:signalregions}The most constraining signal region for each
of the analyses studied in this work. All energies are in units of GeV and
luminosity in fb$^{-1}$.}
\end{center}
\end{table}

\begin{figure}[h!]
\centering
\subfigure{
\includegraphics[width=0.35\textwidth]{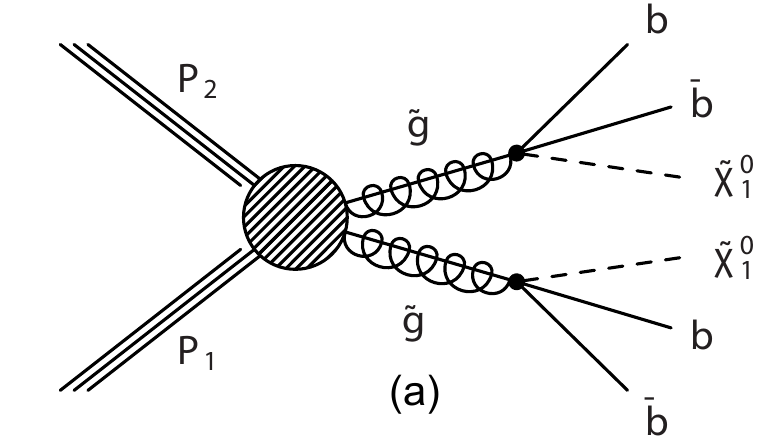}
\label{fig:FG-1}
}
\subfigure{
\includegraphics[width=0.35\textwidth]{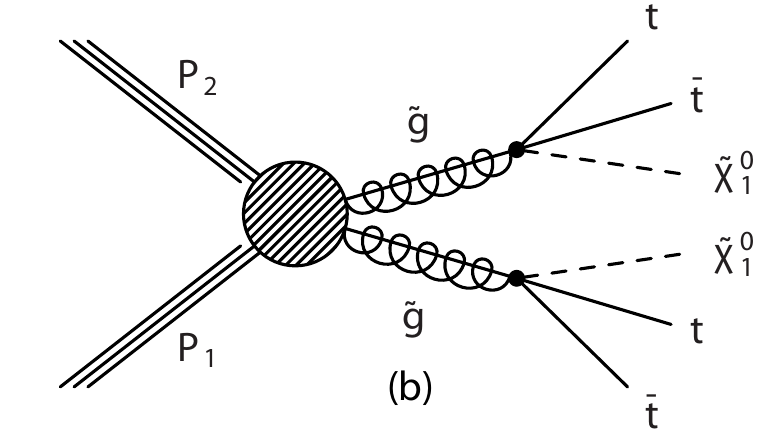}
\label{fig:FG-2}
}
\caption{\footnotesize Feynman diagrams for the simplified SUSY scenarios (taken from Ref. \cite{Chatrchyan:2013wxa}),
T1bbbb (left) and T1tttt (right).
T1bbbb (T1tttt) assumes a gluino branching ratio of 100\% to $b \
\bar{b} \ \tilde{\chi}^0_1 $ ($ t \ \bar{t} \ \tilde{\chi}^0_1$).}
\label{feyndiagrams}
\end{figure}

%%%%%%%%%%%%%%% PLOT 2 %%%%%%%%%%%%%%%
\begin{figure}[h!]
\thisfloatpagestyle{empty}
\centering
\includegraphics[width=0.8\textwidth, trim=0ex 0ex 0ex 3ex, clip]{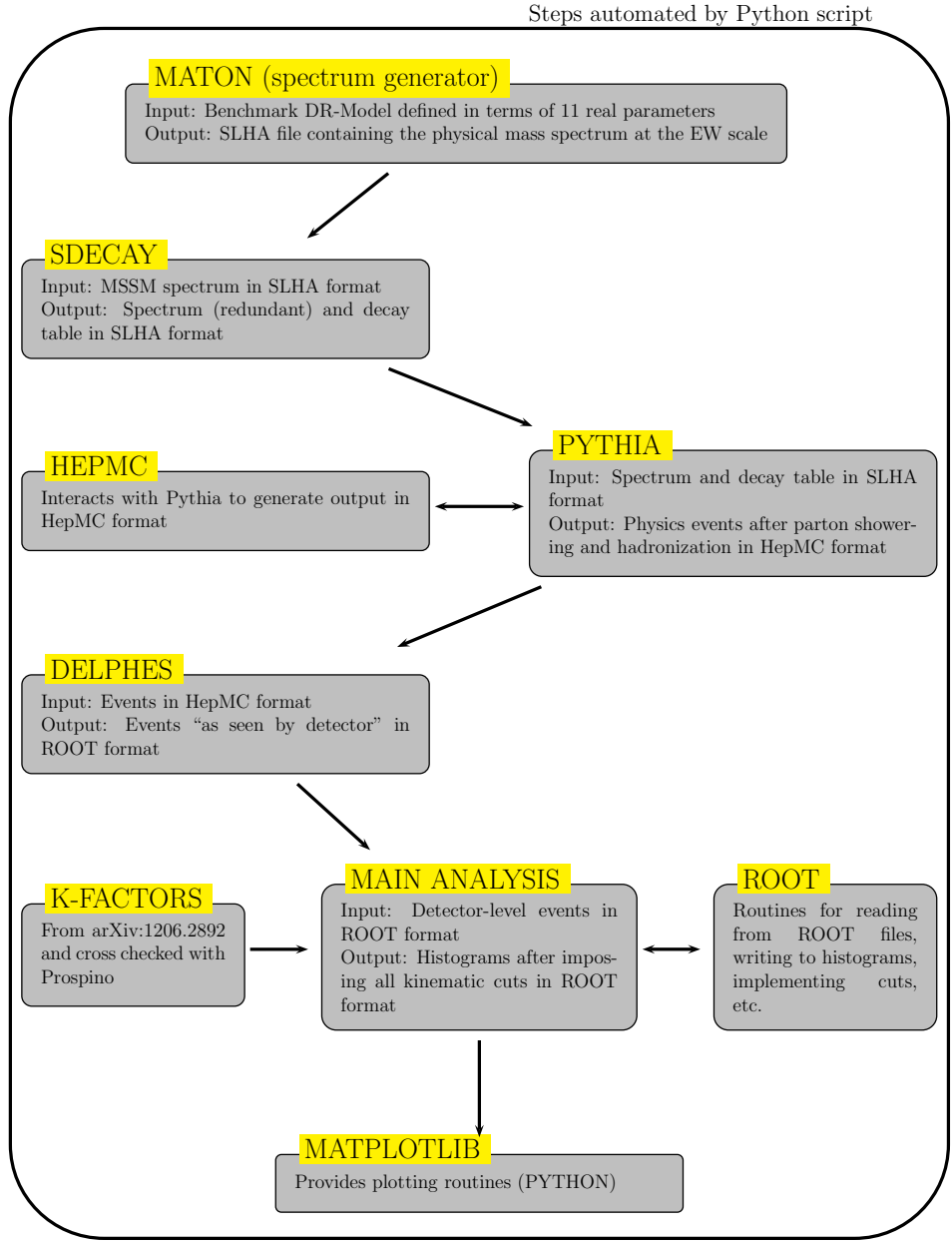}
\caption{\footnotesize The flowchart illustrates the analysis chain for each
benchmark model. The K-Factors are gotten from Ref.~\cite{Kramer:2012bx}. }
\label{fig:flowchart}
\end{figure}
%%%%%%%%%%%%%%%%%%%%%%%%%%%%%%%%%%%%%%

Next we use the the output from \code{maton} as input for the program
\code{SDECAY 1.3} \cite{Muhlleitner:2003vg} which calculates the decay widths
and branching ratios for all the SUSY particles in the spectrum.\footnote{The gluino decay
rates at large $\tan\beta$ were calculated in Refs. \cite{Baer:1990sc,Baer:1997yi,Baer:1998bj}.}  The output is
also in SLHA format and contains along with the decay table also the spectrum
that was calculated in the previous step, because this information is needed for
the next steps in the analysis.

For event generation, we use \code{PYTHIA 8.175} \cite{Sjostrand:2007gs}. We
have written a main program for \code{PYTHIA} that reads in all its parameters
from a so-called ``card file'' and writes the events to a file in \code{HepMC}
format\footnote{We are using \code{HepMC 2.06.09}.} \cite{Dobbs:2001ck}. Our
card file is identical with \filename{main24.cmnd} that comes with \code{PYTHIA}
as a template with the following changes indicated in Listing~\ref{lst:mymain_cmnd}. \\[10pt]

\begin{lstlisting}[label=lst:mymain_cmnd,caption=pythia8175/examples/mymain.cmnd
]
Main:numberOfEvents = 10000       ! number of events to generate
Main:timesAllowErrors = 3         ! how many aborts before run stops

Beams:eCM = 8000.                 ! CM energy of collision

SLHA:file = sps-so10.spc

SUSY:all = off                    ! Switches on ALL (~400)
SUSY:gg2gluinogluino  = on
SUSY:qqbar2gluinogluino  = on

PartonLevel:MPI = on              ! no multiparton interaction
PartonLevel:ISR = on              ! no initial-state radiation
PartonLevel:FSR = on              ! no final-state radiation
HadronLevel:Hadronize = on        ! no hadronization
\end{lstlisting}

The card file is self-explanatory, and we just mention that we generate 10,000
events (line 1), and in line 6 of \ref{lst:mymain_cmnd} we are reading
in the decay file that we had generated in the previous step. Also, note that we
have switched off all SUSY processes {\em except} $gg\ra \widetilde{g}\,
\widetilde{g}$ and $q\bar{q}\ra \widetilde{g}\, \widetilde{g}$, since the only
light particles in the spectrum are the gauginos, and the neutralinos and
charginos will not contribute to the event signatures that we will later
consider for the analyses. Moreover, the electroweak processes have much smaller
cross sections and can also be neglected.

For the detector simulation, we use \code{Delphes 3.0.9}\cite{Ovyn:2009tx}.
Compiling and running the code is straightforward, and the only file that needs
to be adapted to the respective analysis is the \code{Delphes} card that chooses the
detector we want to simulate (ATLAS or CMS) and sets various parameters like the
b-tagging efficiency, etc. Since all these settings depend on the analysis under
consideration, we will give the details in the corresponding sections further
below. From the output of \code{PYTHIA}, \code{Delphes} produces a file that contains
the events ``as seen by the detector'', i.e.~particle tracks, transverse
momentum $p_T$, missing transverse energy $\cancel{E}_T$, etc. The result is
saved as a \code{ROOT} tree \cite{Brun:1997pa}.

We now come to the main part of our analysis. Our dedicated C++ code reads in
the \code{ROOT} file and implements the event selection\footnote{We are using
\code{ROOT\,v5-28-00-patches@42209}.} for the respective analysis. We generate
the C++ classes necessary to read in the events from the \code{ROOT} file
automatically\footnote{We should mention that to obtain a program that compiles,
one needs to include the files TRef.h and TRefArray.h in the header file
that is automatically created by Listing~\ref{lst:create_Analysis_class_C} by
hand; this seems to be a bug in \code{ROOT\,v5-28}.} by the lines given in
Listing \ref{lst:create_Analysis_class_C} (to be run in \code{ROOT}).

\begin{lstlisting}[label=lst:create_Analysis_class_C,
caption=root/SO10-pheno/create\_Analysis\_class.C]
{
  TFile *_file0 = TFile::Open("mymain.root");
  Delphes->MakeClass("Analysis");
}
\end{lstlisting}

The main program implements the cuts and depends on the specific analysis at
hand, and we will give some more details in the following sections. The output
of the analysis (number of events passing the cuts, cross sections, histograms,
etc.) is stored in the form of text and ROOT files. Note that for comparing our
events to experiment, we need to rescale the number of events to the luminosity
of the respective analysis and correct for the leading-order gluino production
cross section in \code{PYTHIA} by using $K$-factors that we obtained from Ref.~\cite{Kramer:2012bx}. We have
compared these $K$-factors with what we get from \code{Prospino
2.1.}\cite{Beenakker:1996ed}, and as expected we find excellent agreement.
Finally, we use
\code{matplotlib 1.0.1.} \cite{Hunter:2007} and \code{Python 2.7.1.} to
visualize the results.

%\clearpage
\section{Same Sign Di-lepton Analysis}
\label{sec:dilepton}

We consider the 15 benchmark models corresponding to different gluino masses
(\ref{tab:benchmark-spectrum} lists 6 of them) and investigate their discovery
potential at LHC. We start with Ref.~\cite{Chatrchyan:2012paa} that performs a
search for new physics in same-sign di-lepton events with at least 2 b-jets, and
concentrate on the signal region 3 (SR3) where the SM background is
lowest\footnote{We have checked that the constraints on our models from the
other signal regions are weaker.}. For the cuts defining SR3, see
\vref{tab:signalregions}.

The parameters regarding the event generation (\code{PYTHIA}) are unchanged. For
the detector simulation, however, we need to adapt the Delphes card to the
current analysis. In particular, in the card file
\filename{delphes\_card\_CMS.tcl} we replace the default b-tagging efficiency by
the one given in Ref.~\cite{Chatrchyan:2012paa}, p.~6.

In the main analysis, we implement all cuts and other event selection criteria
that correspond to SR3 and that are described in detail in
Ref.~\cite{Chatrchyan:2012paa}. We store the number of events passing all cuts,
the total gluino production cross section, and its error in a text file. Before
we can plot these events and compare them to the expected SM background, we need
to rescale our 10,000 gluino events to the integrated luminosity of 10.5
fb${}^{-1}$ of events underlying the current analysis.

\begin{figure}[h]
\centering
\subfigure[Validation.]{
\includegraphics[width=0.45\textwidth, trim=0ex 2ex 0ex 4ex,
clip]{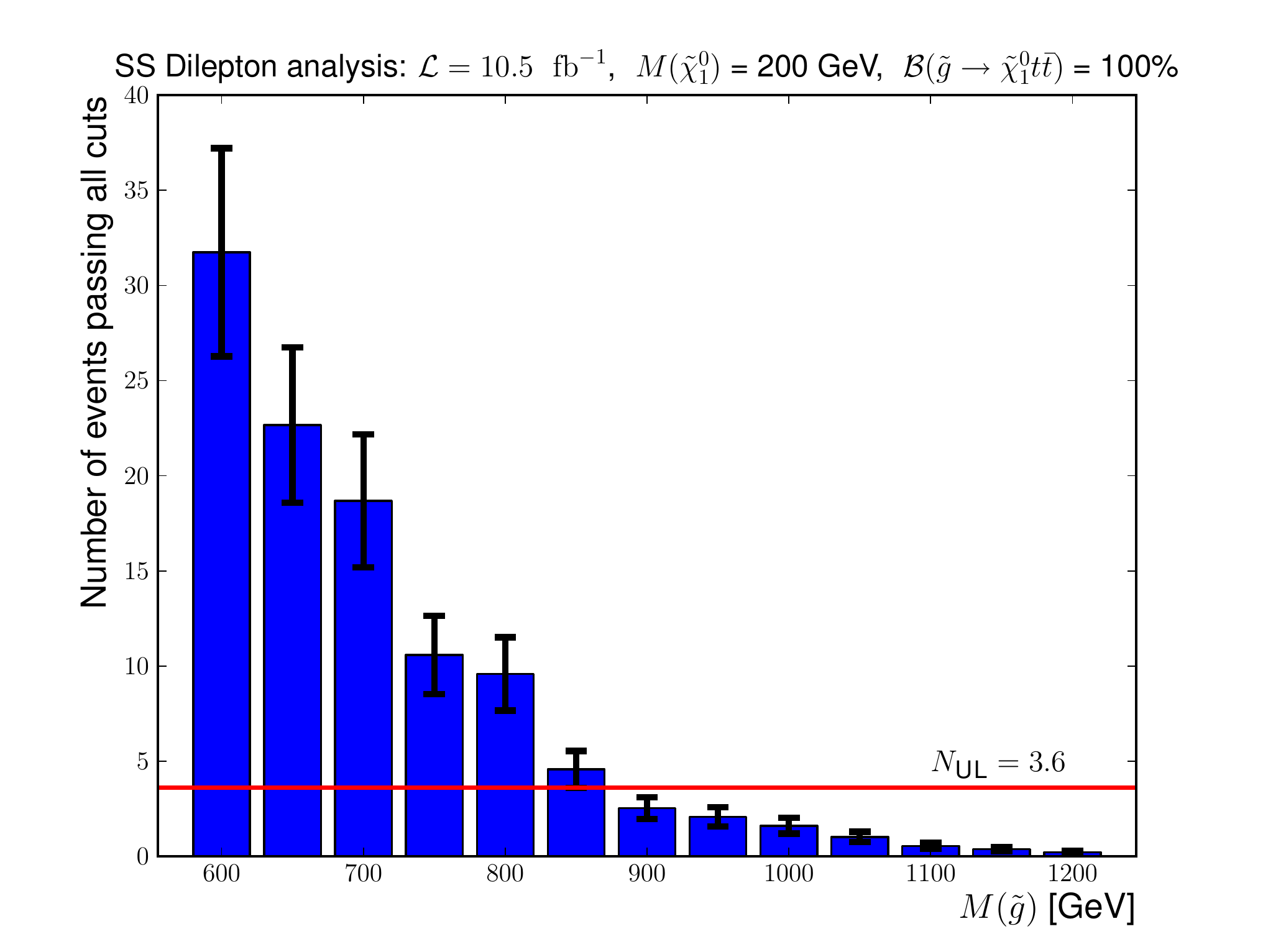}
%\label{fig:FG-1}
\label{fig:validation-CMS-dilepton-neut200}
}
\subfigure[Benchmark]{
\includegraphics[width=0.45\textwidth, trim=0ex 6ex 0ex 4ex,
clip]{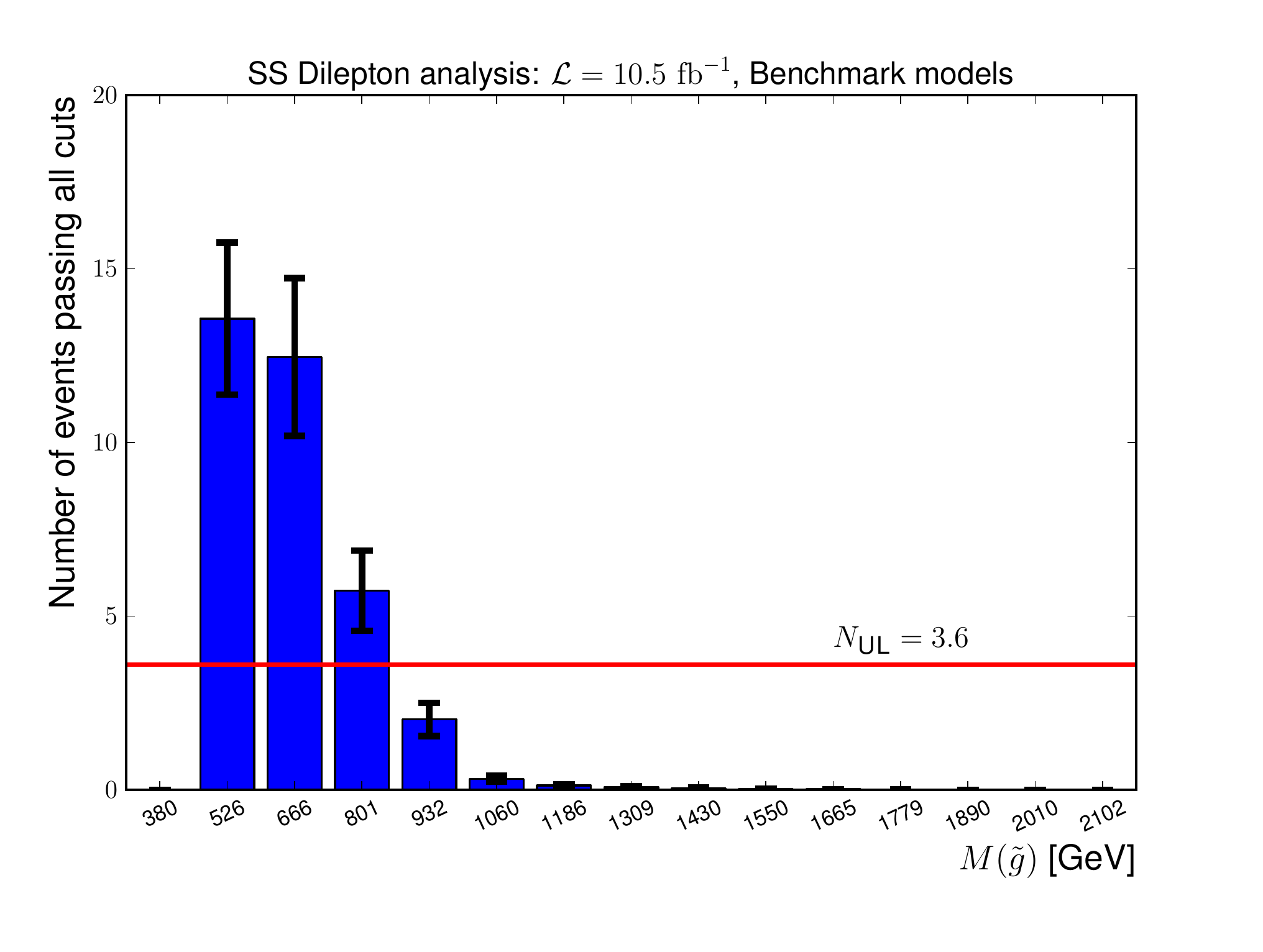}
%\label{fig:FG-1}
\label{fig:CMS-dilepton-benchmark}
}
\caption{\footnotesize The number of events passing all cuts corresponding to
signal region 3 (SR3) for the simplified model A1 \cite{Chatrchyan:2012paa} for
fixed $M_{\widetilde{\chi}^0_1}=200$ GeV (left panel) and for the benchmark
models (right panel). The label on the $x$-axis denotes the gluino
mass $M_{\widetilde{g}}$. The red horizontal line shows the
upper limit on events from new physics at 95\% CL, and the error bars are derived from the uncertainty in
the $K$-factors.}
\label{fig:CMS-dilepton}
\end{figure}

To set limits on the neutralino and gluino masses,
Ref.~\cite{Chatrchyan:2012paa} assumes a set of simplified models. The one that
comes closest to our case is Model A1 (cmp.~\vref{fig:FG-2}). To validate our
analysis, i.e.~to check whether our analysis can reproduce the experimental one
within reasonable error margins, we retraced the steps leading to Fig.~4 of
Ref.~\cite{Chatrchyan:2012paa}. To that end, we started out with a template SLHA
file, set $\mathcal{B}(\widetilde{g}\rightarrow
\widetilde{\chi}^0_1 t\bar{t})=100\%$ and scanned over a range of gluino and
neutralino masses. We exclude a model (given by a pair of gluino and neutralino
masses), if the event count is greater than the 95\% CL upper limit on events from new physics (horizontal red
line). In \ref{fig:validation-CMS-dilepton-neut200}, we present a representative
plot corresponding to a constant $M_{\widetilde{\chi}^0_1}=200$ GeV and read off
that values of $M_{\widetilde{g}} < 875 \pm 25$ GeV are excluded. The agreement
between \ref{fig:validation-CMS-dilepton-neut200} and Fig.~4 of
Ref.~\cite{Chatrchyan:2012paa} (which gives a lower limit on the gluino mass of $\approx 1000 \pm 30$ GeV) is well within 20\%, which is the expected
precision that can be achieved discounting for the fact that we do not have
access to the full array of experimental tools used by the ATLAS and CMS
collaborations (in particular the detector simulation).
In \ref{fig:CMS-dilepton-benchmark} we present the result of our analysis for
the benchmark models. As is clear from the plot, models with $M_{\tilde g}
\gtrsim850$ GeV yield event numbers that are below the 95\% upper limit allowed from
models of new physics and are
hence not excluded. Note that SR3 is not sensitive to very small gluino masses,
but these are nevertheless excluded by other signal regions. We conclude that in
this case, the simplified model captures the main features of the benchmark
models and thus leads to similar collider bounds or from
Ref.~\cite{Chatrchyan:2012paa} we have $M_{\tilde g} > 1000 \pm 30$ GeV.

%\clearpage

\section{Hadronic Analyses}
\label{sec:hadronic}

%%%%%%%%%%%%%%%%%%%%%%%%%%%%% HADRONIC CMS %%%%%%%%%%%%%%%%%%%%%%%%%%%%%%%

\subsection{$\boldsymbol{\alpha_T}$ Analysis}

The CMS search in Ref. \cite{Chatrchyan:2013lya} looks for final states with
multiple b-jets and missing transverse energy
using a parameter $\a_T$, on which we briefly elaborate below, and interprets
the results in the context of simplified
models where one assumes only one decay channel, respectively, given by the
Feynman diagrams in \ref{feyndiagrams}. The cuts in this search require large
$E_T$
and $H_T$ for the jets, no isolated electrons, muons, or photons, and a cut on
$\a_T$. This search was performed at $\sqrt{s} = 8$ TeV and
used 11.7 $\text{fb}^{-1}$ of data. The analysis considers only jets with
$E_T > 50$ GeV and $|\eta| < 3.0$ and bins the
events according to $H_T$ and the number of jets and b-jets. Binning details can
be found in Tab. 4 of Ref. \cite{Chatrchyan:2013lya}.

The parameter $\a_T$ is used to reject events with either little $\slashed{E}_T$
or a transverse energy mismeasurement. Furthermore, it is sensitive to
final states containing large, genuine $\slashed{E}_T$. In a dijet system,
$\a_T$ is defined as the ratio of the softer jet's transverse energy to the
transverse mass of the dijet system. For events with more than two jets, a
pseudo-dijet system can be formed so that $\a_T$ may still be employed. For
an event in which the two jets are back-to-back, have equal transverse energy,
and have momenta that are large compared to their respective masses,
$\a_T$ has a value of 0.5. In the same limit, if the two jets are back-to-back
but are measured to have different transverse energies, then $\a_T$ will
be less than 0.5. Events in which the two jets are not back-to-back and large,
genuine $\slashed{E}_T$ is present, then $\a_T$ will be much larger than 0.5.
We therefore require $\a_T > 0.55$.
\begin{figure}[h]
\centering
\subfigure{
\includegraphics[width=0.45\textwidth]{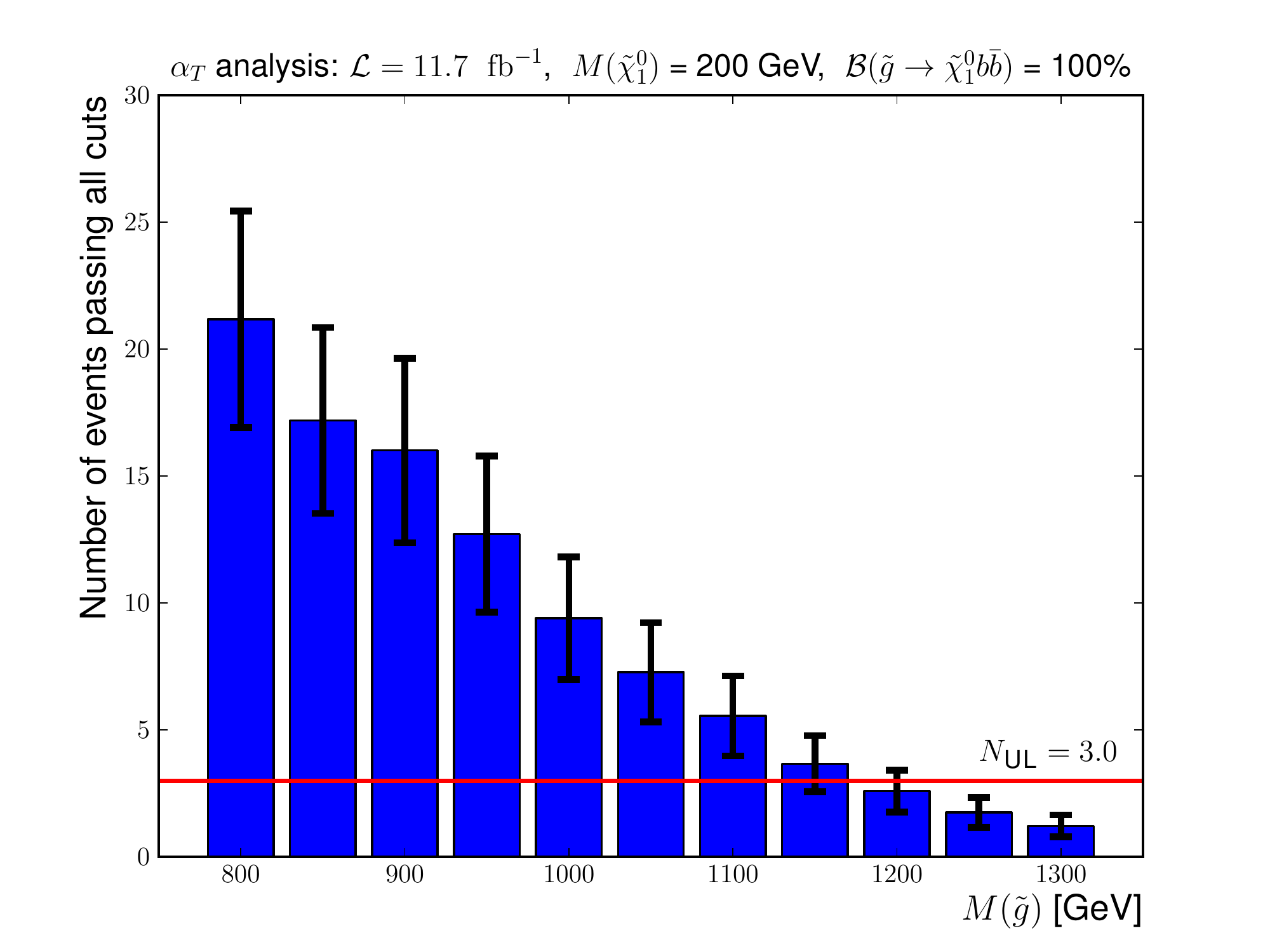}
%\label{fig:FG-1}
}
\subfigure{
\includegraphics[width=0.45\textwidth]{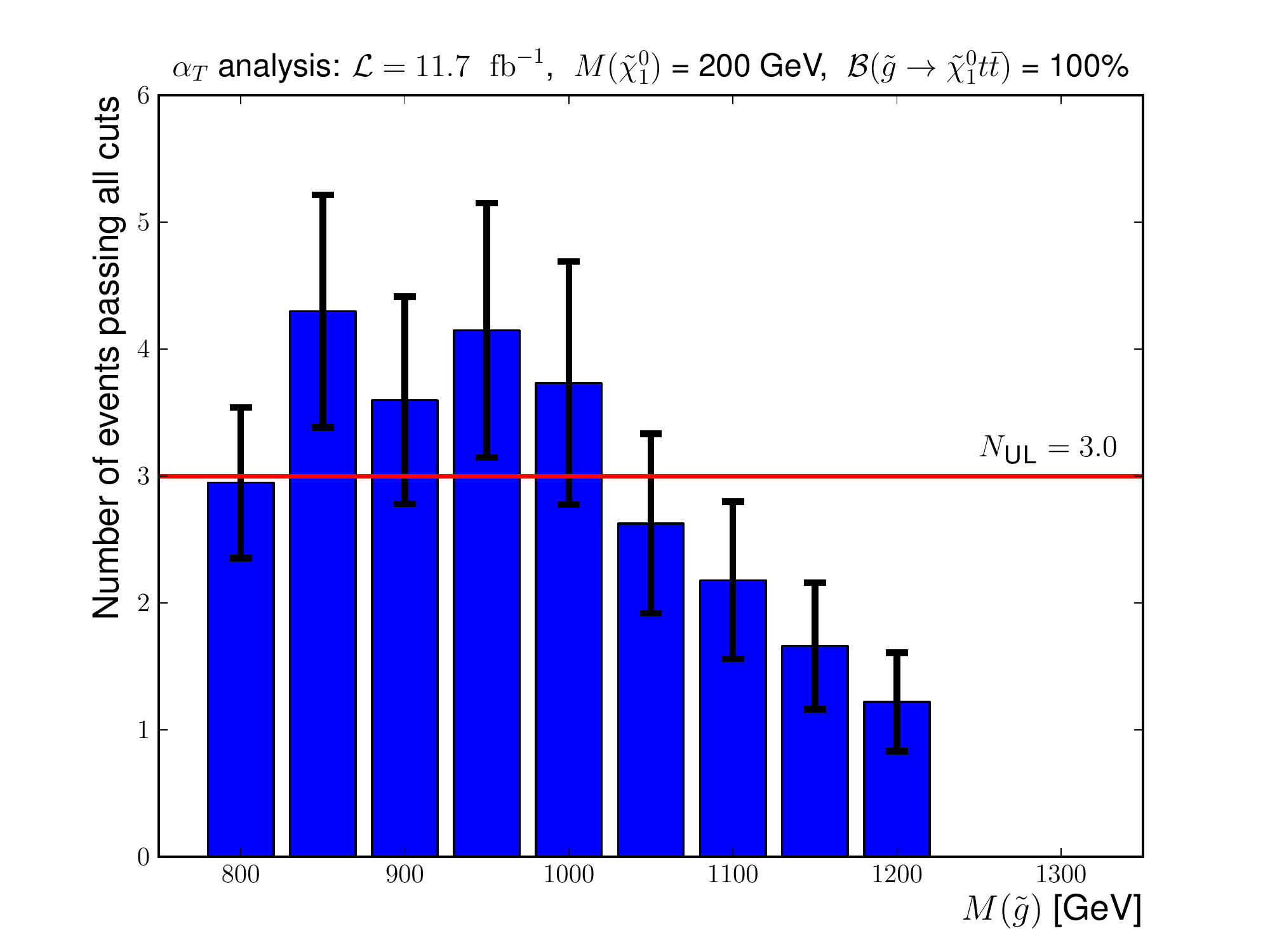}
%\label{fig:FG-1}
}
\caption{\footnotesize Validation of CMS analysis \cite{Chatrchyan:2013lya} for
the simplified SUSY scenarios,
$\tilde{g} \rightarrow  b \ \bar{b} \ \tilde{\chi}^0_1$ (left) and $\tilde{g}
\rightarrow t \ \bar{t} \ \tilde{\chi}^0_1 $ (right).
The CMS analysis rules out gluinos lighter than 1160 GeV in the T1bbbb model and
1025 GeV in the T1tttt simplified models.
Our results are in excellent agreement with the CMS bounds.}
\label{alphaTvalidation}
\end{figure}

The Delphes CMS card is modified to contain the electron, muon, and photon
isolation criteria given in Sec. 4 of Ref. \cite{Chatrchyan:2013lya}. In the same
section,
the paper states that the b-tagging efficiency is 60-70\% and $p_T$-dependent.
We therefore use the same b-tagging efficiency as for the di-lepton analysis. To
validate
our procedure for this analysis, we calculate bounds on the T1bbbb and T1tttt
simplified models (\ref{feyndiagrams}) assuming 100\% branching ratios to
$\tilde{\chi}^0_1b\bar{b}$ and $\tilde{\chi}^0_1t\bar{t}$, respectively, using
the observed number of events and the expected SM background. These bounds are
indicated in \ref{alphaTvalidation} by the horizontal red line annotated by
$N_\text{UL}$ for an LSP of mass 200 GeV, which is the region most relevant for
our model.
The bounds on the gluino mass obtained from our (the CMS) analysis are 1130
(1125) GeV in the T1bbbb model and 975 (950) GeV in the T1tttt model. The values
obtained
in the validation are in excellent agreement with those in the CMS analysis
\cite{Chatrchyan:2013lya}.

\begin{figure}[h]
\thisfloatpagestyle{empty}
\centering
\includegraphics[width=0.5\textwidth]{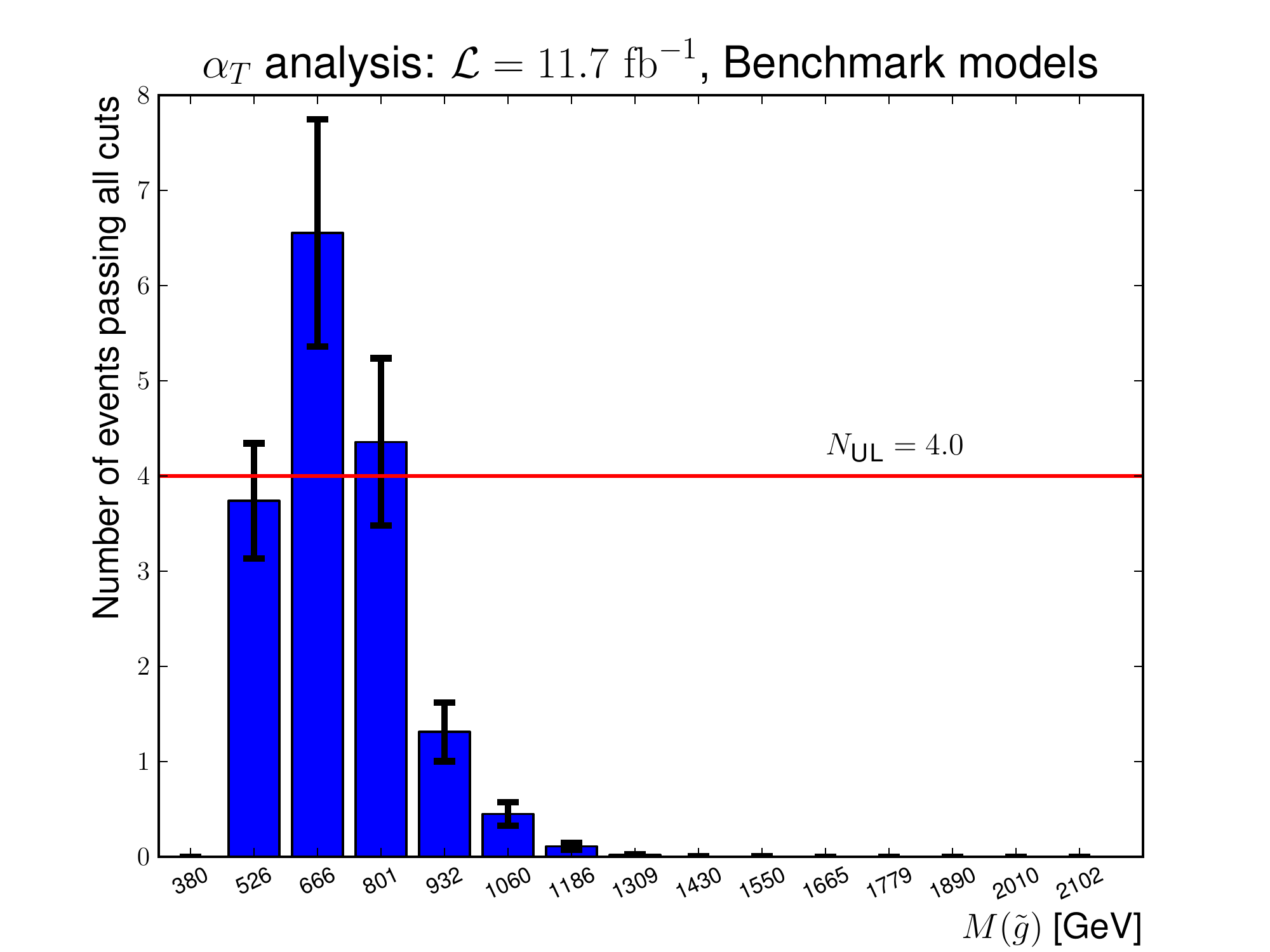}
\caption{\footnotesize The number of events passing all cuts corresponding to
the bin with 4 or more jets, 2 b-jets, and $775 < H_T (\text{GeV}) < 875$ in
Ref.\cite{Chatrchyan:2013lya} for the benchmark models. The labels on the
$x$-axis denote the gluino mass in each benchmark model. The red
horizontal line shows the  95\% upper limit on the allowed events from new
physics. The error bars are derived from the uncertainty in estimating
the NLO gluino production cross-section.}\label{fig:CMS-alphaT}
\end{figure}

\newpage
We find the bins described in \ref{tab:signalregions} of this paper to be the most
constraining. Note that these bins select a particular number of b-jets.
In the benchmark models, the gluinos have a significant branching ratio to a
gluon and a neutralino. Since this decay mode does not result in b-jets,
the benchmark models are not well approximated by the simplified models for this
analysis. We typically find that the benchmark models give fewer b-jets.
Thus the simplified models and the benchmark models do not receive the strongest
constraints from the same bin.

After successfully validating our analysis, we now move on to analyze our
benchmark models. The number of events that pass the cuts for each
benchmark model is shown in \ref{fig:CMS-alphaT}. The bounds are clearly much
weaker for the benchmark models. While the T1bbbb (T1tttt) simplified models
rule out gluinos with mass 1125 (950) GeV, we rule out only gluinos with mass
$\approx 775$ GeV for this analysis.

%\clearpage
\subsection{$\Delta \hat{\phi}$ Analysis}
In Ref. \cite{Chatrchyan:2013wxa}, the CMS collaboration performed a search for
gluinos decaying according to one of the simplified SUSY scenarios shown in
\ref{feyndiagrams} by looking for events with large transverse missing energy,
jets and b-jets, and with no isolated electrons or muons. The data sample
used in this analysis was recorded at 8 TeV center of mass energy and includes
19.4 fb$^{-1}$ of data. The analysis required at least 3 jets (and at least 1
b-tagged jet)
with $p_T > 50$ GeV and $|\eta| < 2.4$, and binned the events into 14 signal
regions with different ranges of $\slashed{E}_T$, $H_T$ and
$N_{\text{b-jet}}$.
The details of the binning can be found in Tab. 1 and Tab. 2 of Ref.
\cite{Chatrchyan:2013wxa}. For neutralino masses of 0 - 400 GeV, we find the
signal region summarized in the last row of \ref{tab:signalregions} to be the
most constraining one. In addition, events are required to have
$\Delta \hat{\phi}_{\text{min}} > 4.0$, where $ \Delta \hat{\phi}_{\text{min}} =
\text{min} \left( \Delta \phi_i/\sigma_{\Delta \phi_i} \right) $
and $\Delta \phi$ is the angle between a jet and the negative of the
$\slashed{E}_T$ vector, and $\sigma_{\Delta \phi_i}$ is the estimated
resolution of $\Delta \phi$. More on this observable can be found in Ref.
\cite{Chatrchyan:2012rg}. QCD background events
are characterized by low $\Delta \phi$, since the $p_T$ mis-measurement gives
rise to most of the missing energy in a QCD event.  By requiring
$\Delta \hat{\phi}_{\text{min}} > 4.0$, most of the QCD backgrounds are
eliminated.

\begin{figure}[h]
\centering
\subfigure{
\includegraphics[width=0.45\textwidth]{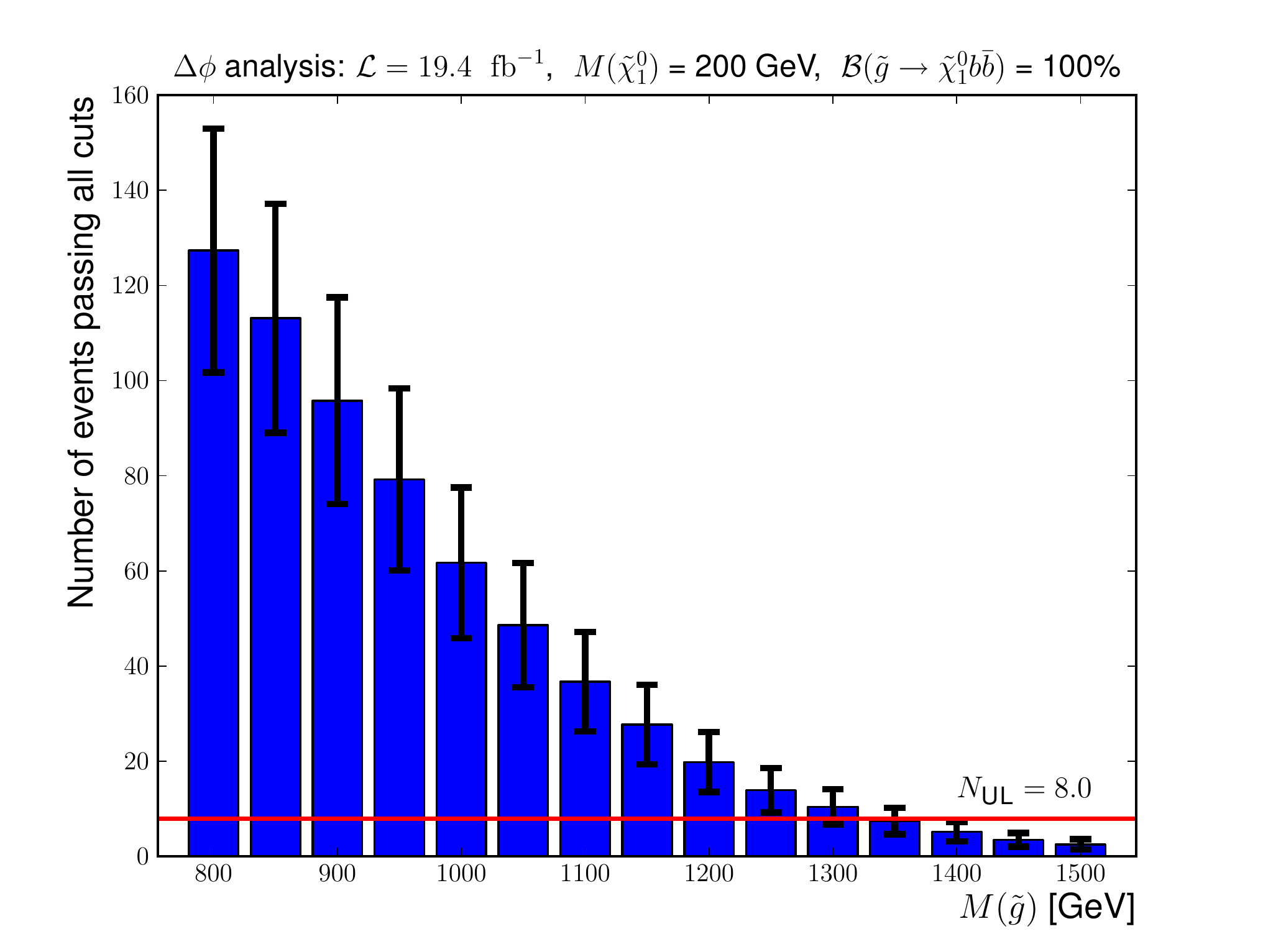}
%\label{fig:FG-1}
}
\subfigure{
\includegraphics[width=0.45\textwidth]{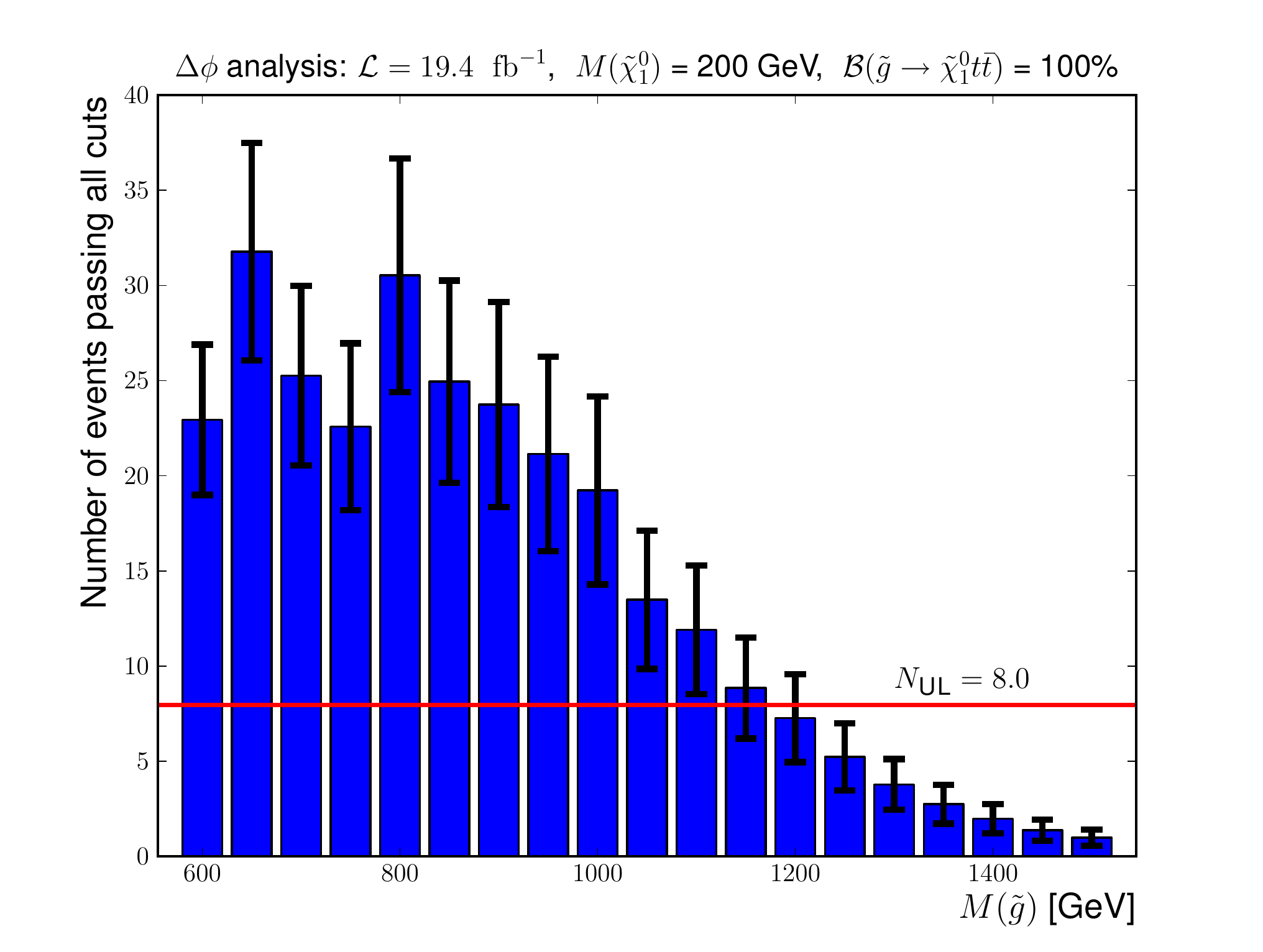}
%\label{fig:FG-1}
}
\caption{\footnotesize Validation of CMS analysis \cite{Chatrchyan:2013wxa} for
the simplified SUSY scenarios, $\tilde{g} \rightarrow \tilde{\chi}^0_1 b \bar{b}
$ (left) and
$\tilde{g} \rightarrow \tilde{\chi}^0_1 t \bar{t} $ (right). The CMS analysis
rules out gluinos lighter than 1170 GeV in the T1bbbb model and
1020 GeV in the T1tttt simplified models. Our results are in agreement with the
CMS bounds within the expectations of our numerical tools.}
\label{delphivalidation}
\end{figure}

As with the previous analyses, we first validate the CMS analysis. For this
analysis, CMS quotes that the nominal
b-tagging efficiency is 75\% for jets with a $p_T$ value of 80 GeV\footnote{We
thank Mike Saelim for a digitized version of the b-tagging efficiency
from Ref. \cite{Chatrchyan:2012jua}.}. We adapt the Delphes CMS card using the
following information from the CMS analysis: b-tagging efficiency, electron,
and muon isolation criteria, and charged track definition and isolation
criteria. Using the information on the number of observed events and the
expected Standard Model
background from Ref. \cite{Chatrchyan:2013wxa} we determine bounds for the
simplified models T1tttt and T1bbbb. The result is shown in
\ref{delphivalidation}. The bounds on the gluino mass obtained from our validation analysis (CMS
analysis) are
1250 (1170) GeV in the T1bbbb model and 1100 (1020) GeV in the T1tttt simplified
model. The agreement between the two quoted numbers is within
the expectations from the tools that we have used.

\begin{figure}[h!]
\thisfloatpagestyle{empty}
\centering
\includegraphics[width=0.555\textwidth]{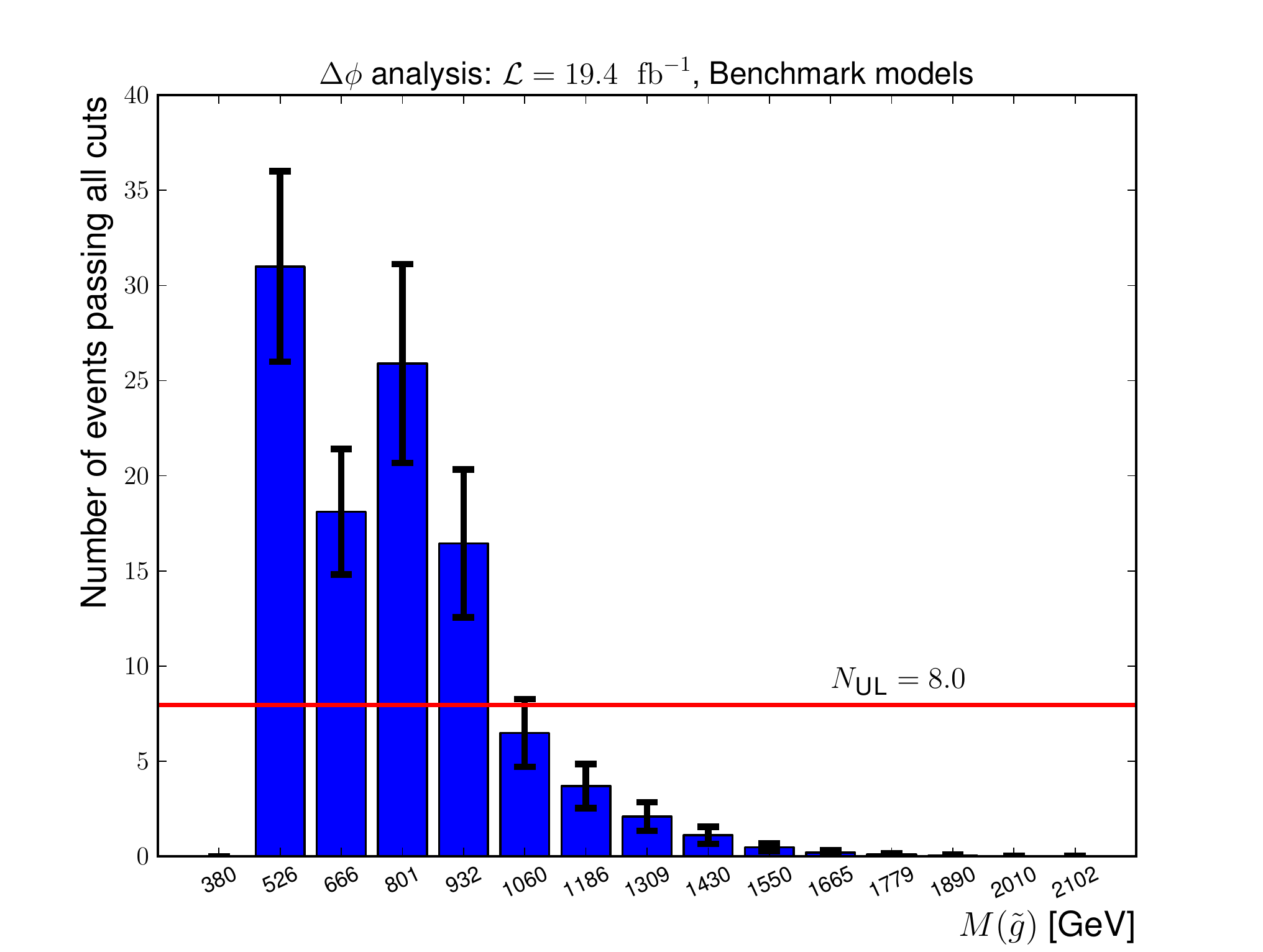}
\caption{\footnotesize The number of events passing all cuts described in
\ref{tab:signalregions} for the benchmark models.
The labels on the $x$-axis denote the gluino mass in each benchmark model. The
red horizontal line shows the  95\% upper limit on the allowed events
from new physics. The error bars are derived from the uncertainty in estimating
the NLO gluino production cross-section.}
\label{fig:CMS-delphi}
\end{figure}

We now proceed to show the results of this analysis, interpreted for the
benchmark models discussed in \ref{sec:benchmark}. We generate 10,000 events for
each of the
benchmark models and apply the same cuts, and since these benchmark models do
not fall into either of the simplified scenarios, we expect the bounds
on the benchmark models to be weaker. In many cases, especially when the mass of
the gluino is greater than 1 TeV, loop decays to a gluon and a neutralino are
enhanced and the final states will not have b-tagged jets. The number of events that pass the cuts for
each benchmark model is shown in \ref{fig:CMS-delphi}. It is clear from the
figure that fewer events pass the cuts, since the final states from the gluino decays
do not come from a single decay topology and there are, on average, fewer b-jets.
Thus, in comparison with the simplified models, where we were able to rule out gluinos of 1100 GeV, we rule out gluinos only
lighter than $\approx 1000$ GeV with this analysis.

%%%%%%%%%%%%%%%%%%%%%%%%%%%%% CONCLUSIONS %%%%%%%%%%%%%%%%%%%%%%%%%%%%%%%

\section{Conclusions}
\label{sec:conclusions}

In this paper, we studied the LHC phenomenology of Yukawa unified \SO{10} models. From a wide range of available experimental analyses, we picked those that are most applicable to our case: The same sign di-lepton analysis \cite{Chatrchyan:2012paa} and two hadronic analyses, namely $\alpha_T$ \cite{Chatrchyan:2013lya} and $\Delta \hat{\phi}$ \cite{Chatrchyan:2013wxa}. These analyses are based on simplified models with typically only one decay channel for the gluino, and we re-interpreted their results to set limits on Yukawa unified \SO{10} models.

The first question we addressed was: Which models do survive? To that end, we picked 15 benchmark models strategically placed along the $\chi^2$ curve that constitutes the best fit of Yukawa unified \SO{10} models to low-energy data (cf.~\ref{fig:CL-intervals} and \ref{tab:benchmark-spectrum}), and then looked at the signal events that pass all cuts for the respective analysis. We found that most of the benchmark models with a gluino mass of approximately 1000 GeV or greater survive and cannot be ruled out with current data (cf.~\ref{fig:CMS-dilepton-benchmark}, \ref{fig:CMS-alphaT}, \ref{fig:CMS-delphi}).  Hence many of our benchmark points, \ref{tab:benchmark-spectrum}, are not ruled out by the present LHC data.  In particular,  benchmark points YUc - YUf (\ref{tab:benchmark-spectrum}) are still viable models which can be tested at LHC 14.  As an important side remark, we mention that before analyzing the benchmark models, we validated our analysis by reproducing the results for the simplified models. In all cases, the agreement was within 20\%, and by common lore, this is as close as we might expect to get to the results produced by the experimental collaborations with their more elaborate tools. This constitutes a non-trivial cross-check of our analysis procedure outlined in \ref{fig:flowchart}.

Another question we addressed was the dichotomy between benchmark and simplified models: How well do simplified models describe the Yukawa unified ones? Simplified models are designed to capture the main features of specific scenarios at the cost of (possibly over-)simplifying assumptions. Our findings indicate that for the same sign dilepton analysis, the simplified models describe our benchmark scanarios pretty well, but for the two hadronic analyses, the agreement is less pronounced. The reason is that for the benchmark models, there is a sizeable  branching fraction for loop decays of the gluino to a gluon and a neutralino. However, the simplified models studied so far assume that all decays of the gluino are accompanied by b-quarks. As a consequence, the limits on the gluino mass that we obtain for the benchmark models are weaker than those in the case of the simplified scenarios.

One could argue that the bounds on the gluino mass obtained from simplified and benchmark models differ by less than 20\% and are thus within the expected precision. However, the excellent agreement between the gluino bounds from Refs.~\cite{Chatrchyan:2013lya,Chatrchyan:2013wxa} and our validation (cf.~\ref{alphaTvalidation} and \ref{delphivalidation}) encourages us to believe that this is indeed a physical effect and not a relic of our analysis procedure. In other words, for the hadronic analyses, the agreement in the validation of the benchmark models is about 8 - 10\%, whereas the
bounds we obtain for the benchmark models from the same analysis is less by about 20\%. That is why we conclude that our gluino bounds are weaker.

Lastly, we would like to point out that the $\chi^2$ analysis of Yukawa unified
\SO{10} models suggests an upper bound on the gluino mass of about 3 TeV (for
$m_{16}\lesssim 30$ TeV). Since the amount of fine-tuning increases with
$m_{16}$, larger values of the gluino mass are theoretically disfavored.  In
addition, in our previous paper, Ref. \cite{Anandakrishnan:2012tj}, we showed
that $\chi^2$ was minimized for both the three family and third family analysis
at $m_{16} = 20$ TeV with an upper bound $m_{\tilde g} \leq 2$ TeV (see also
\ref{fig:CL-intervals}).  For larger $m_{16}$, the pull to $\chi^2$ for $M_W$
increased.\footnote{A similar result was found in Ref. \cite{Barger:2012hr}.}
The LHC, after collecting 300 fb$^{-1}$ of data at 14 TeV, has a 5 $\sigma$ discovery potential
for gluinos with mass $\lesssim 1.9$ TeV \cite{Baer:2012vr,CMS:2013xfa}. Thus in
the most favored range of parameters the gluino should be observed at the LHC 14
TeV.  Finally, Yukawa unified models make other significant predictions for the
SUSY particle spectrum as can be seen for our benchmark models
(\ref{tab:benchmark-spectrum}).

%%%%%%%%%%%%%%%%%%% ACKNOWLEDGMENTS %%%%%%%%%%%%%%%%%%%

\section*{Acknowledgments}

We acknowledge useful discussions with Sabine Kraml (Grenoble), Sezen Sekmen (Florida State), Andre Lessa (Sao Paulo), Joerg Meyer (Karlsruhe), Noel Dawe (Vancouver/CERN), Peter Waller
(Manchester) and Radovan Dermisek (Indiana). A.A.~and S.R.~received partial support for this work from
DOE/ER/01545-90. A.A.~and A.W.~acknowledge partial support from LabEx ENIGMASS,
and A.A.~would like to thank the LPSC Grenoble where part of this work was
completed for their hospitality. We thank the \emph{Ohio Supercomputer Center}
and the \emph{Centre de Calcul de l'Institut National de Physique Nucl\'{e}aire
et Physique des Particules} in Lyon for using their resources.  S.R. thanks the Galileo Galilei Institute for Theoretical Physics for the hospitality and the INFN for partial support and CETUP 2013 for their hospitality and partial support during the completion of this work.  S.R. would also like to thank CETUP* (Center for Theoretical Underground Physics and Related Areas), supported by the US Department of Energy under Grant No. DE-SC0010137 and by the US National Science Foundation under Grant No. PHY-1342611, for its hospitality and partial support during the 2013 Summer Program.

%\clearpage

\newpage

\appendix

\labelformat{section}{Appendix #1}
\section{Upper bound on gluino mass}
\label{app:gluino}
\begin{figure}[h!]
\thisfloatpagestyle{empty}
\centering
\includegraphics[width=0.8\textwidth, trim=0ex 0ex 0ex 4ex,
clip]{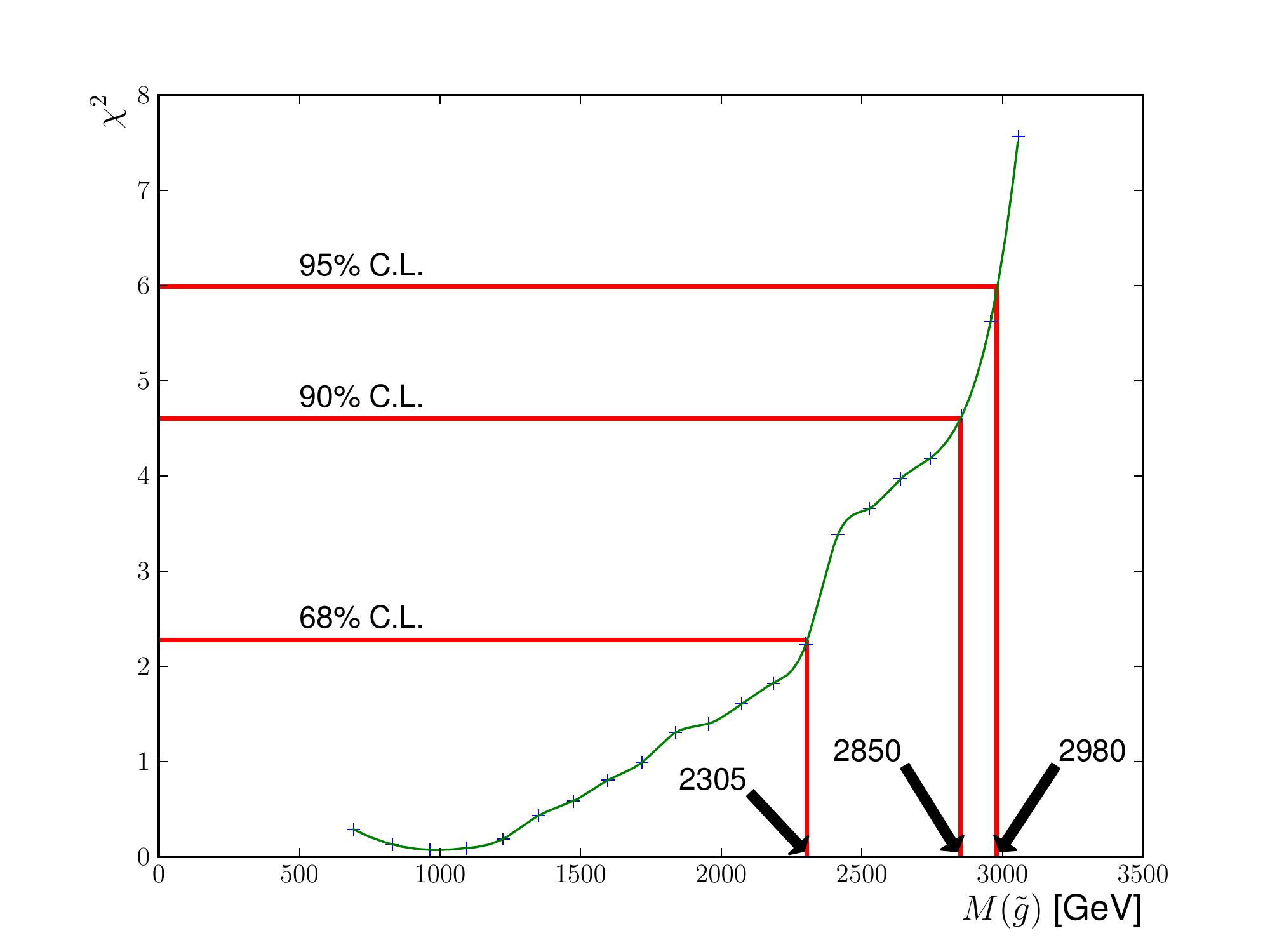}
\caption{\footnotesize The $\chi^2$ function for fixed $m_{16} = 30$ TeV and
different values of $M_{1/2}$ corresponding to the gluino masses indicated on
the $x$-axis. In minimizing $\chi^2$ for each point, we followed the same
procedure as in Ref.~\cite{Anandakrishnan:2012tj} except that we now do not
penalize for gluino masses smaller than a certain lower bound. The confidence
level intervals correspond to the $\chi^2$ distribution with two degrees of
freedom.  Note, the 90\% upper bound on the gluino mass increases as $m_{16}$ increases.}
\label{fig:30TeV-confidencelevels}
\end{figure}

%%%%%%%%%%%%%%%%%%%%%%%%%%%%%%%%%%%%%%%%%%%%%%%%%%%%%%%%%%%
Note, in \ref{fig:30TeV-confidencelevels} we take $m_{16} = 30$ TeV. We find that the upper bound on the gluino mass increases as $m_{16}$ increases.
This is due to the combination of the constraints from the bottom quark and light Higgs mass.  For $m_{16} = 30$
TeV the upper bound on the gluino mass at 90\% CL is 2.85 TeV.  Although in our three family analysis $\chi^2$ increases as $m_{16}$ increases,
we do not believe that there is a strict upper bound on $m_{16}$ other than the fact that the level of fine-tuning increases as $m_{16}$ increases.

%%%%%%%%%%%%%%%%%%%%%%%%%%%%% APPENDIX: BENCHMARK MODELS
%%%%%%%%%%%%%%%%%%%%%%%%%%%%%%%
\newpage
\section{Details on the Benchmark Models}
\label{app:benchmark}

\begin{table}[h!]
\begin{center}
\begin{footnotesize}
 \caption[8]{ \mbox{{\bf Benchmark model YUa}}\\ (1/$\alpha_G, \, M_G, \, \epsilon_3, \ \lambda$) = ($25.95,\, 2.87 \times 10^{16}$ GeV, $\, -1.04$ \%, $0.603$),\\
($m_{16}, \, M_{1/2}, \, A_0, \, \mu(M_Z)$) = ($20000,\, 150, \,
-41289, \, 869$) GeV,\\
($(m_{H_d}/m_{16})^2, \, (m_{H_u}/m_{16})^2, \, \tan\beta$) = ($1.87,  \, 1.62, \, 49.69$)
\label{t:YUa} % end of caption
}
\end{footnotesize}
\begin{tabular}{|l|c|c|c|c|}
\hline
Observable & Fit value & Experimental Value & Pull & Error \\
\hline
$M_Z$ &              91.1876         &  91.1876         &  0.0000          &  0.4559          \\
$M_W$ &              80.5166         &  80.4360         &  0.2004          &  0.4025          \\
$1/\alpha_{em}$ &    137.1232        &  137.0360        &  0.1272          &  0.0000          \\
$G_{\mu} \times 10^5$ & 1.1689          &  1.1664          &  0.2172          &  0.0117          \\
$\alpha_3$ &         0.1184          &  0.1184          &  0.0337          &  0.0009          \\
\hline
$M_t$ &              173.5973        &  173.5000        &  0.0735          &  1.3238          \\
$m_b(m_b)$ &         4.1815          &  4.1800          &  0.0403          &  0.0366          \\
$m_{\tau}$ &         1.7766          &  1.7768          &  0.0242          &  0.0089          \\
\hline
$M_h$ &              124.33          &  125.30          &  0.3168          &  3.0676          \\
\hline
$\mathcal{B}(B \rightarrow X_s \gamma) \times 10^4$ & 3.2670          &  3.4300          &  0.0995          &  1.6374          \\
$\mathcal{B}(B_s \rightarrow \mu^+ \mu^-) \times 10^9$ & 3.1369          &  3.2000          &  0.0387          &  1.6308          \\
\hline
\multicolumn{3}{|l}{Total $\chi^2$}  & {\bf 0.224014}&  \\\hline
\end{tabular}
\end{center}
\end{table}

\begin{table}[h!]
\begin{center}
\begin{footnotesize}
 \caption[8]{ \mbox{{\bf Benchmark model YUb}}\\ (1/$\alpha_G, \, M_G, \, \epsilon_3, \ \lambda$) = ($26.00,\, 2.71 \times 10^{16}$ GeV, $\, -0.94$ \%, $0.603$),\\
($m_{16}, \, M_{1/2}, \, A_0, \, \mu(M_Z)$) = ($20000,\, 200, \,
-41272, \, 889$) GeV,\\
($(m_{H_d}/m_{16})^2, \, (m_{H_u}/m_{16})^2, \, \tan\beta$) = ($1.87,  \, 1.62, \, 49.69$)
\label{t:YUb} % end of caption
}
\end{footnotesize}
\begin{tabular}{|l|c|c|c|c|}
\hline
Observable & Fit value & Experimental Value & Pull & Error \\
\hline
$M_Z$ &              91.1876         &  91.1876         &  0.0000          &  0.4559          \\
$M_W$ &              80.5133         &  80.4360         &  0.1920          &  0.4025          \\
$1/\alpha_{em}$ &    137.1111        &  137.0360        &  0.1096          &  0.0000          \\
$G_{\mu} \times 10^5$ & 1.1688          &  1.1664          &  0.2047          &  0.0117          \\
$\alpha_3$ &         0.1184          &  0.1184          &  0.0523          &  0.0009          \\
\hline
$M_t$ &              173.5907        &  173.5000        &  0.0685          &  1.3238          \\
$m_b(m_b)$ &         4.1817          &  4.1800          &  0.0478          &  0.0366          \\
$m_{\tau}$ &         1.7766          &  1.7768          &  0.0302          &  0.0089          \\
\hline
$M_h$ &              123.58          &  125.30          &  0.5602          &  3.0676          \\
\hline
$\mathcal{B}(B \rightarrow X_s \gamma) \times 10^4$ & 3.2584          &  3.4300          &  0.1048          &  1.6374          \\
$\mathcal{B}(B_s \rightarrow \mu^+ \mu^-) \times 10^9$ & 3.1384          &  3.2000          &  0.0378          &  1.6308          \\
\hline
\multicolumn{3}{|l}{Total $\chi^2$}  & {\bf 0.427658}&  \\\hline
\end{tabular}
\end{center}
\end{table}

\begin{table}[h!]
\begin{center}
\begin{footnotesize}
 \caption[8]{ \mbox{{\bf Benchmark model YUc}}\\ (1/$\alpha_G, \, M_G, \, \epsilon_3, \ \lambda$) = ($25.97,\, 2.88 \times 10^{16}$ GeV, $\, -1.30$ \%, $0.605$),\\
($m_{16}, \, M_{1/2}, \, A_0, \, \mu(M_Z)$) = ($20000,\, 250, \,
-41231, \, 824$) GeV,\\
($(m_{H_d}/m_{16})^2, \, (m_{H_u}/m_{16})^2, \, \tan\beta$) = ($1.87,  \, 1.62, \, 49.68$)
\label{t:YUc} % end of caption
}
\end{footnotesize}
\begin{tabular}{|l|c|c|c|c|}
\hline
Observable & Fit value & Experimental Value & Pull & Error \\
\hline
$M_Z$ &              91.1876         &  91.1876         &  0.0000          &  0.4559          \\
$M_W$ &              80.5329         &  80.4360         &  0.2407          &  0.4025          \\
$1/\alpha_{em}$ &    137.1585        &  137.0360        &  0.1786          &  0.0000          \\
$G_{\mu} \times 10^5$ & 1.1694          &  1.1664          &  0.2629          &  0.0117          \\
$\alpha_3$ &         0.1183          &  0.1184          &  0.0758          &  0.0009          \\
\hline
$M_t$ &              173.6810        &  173.5000        &  0.1367          &  1.3238          \\
$m_b(m_b)$ &         4.1827          &  4.1800          &  0.0750          &  0.0366          \\
$m_{\tau}$ &         1.7763          &  1.7768          &  0.0532          &  0.0089          \\
\hline
$M_h$ &              123.01          &  125.30          &  0.7474          &  3.0676          \\
\hline
$\mathcal{B}(B \rightarrow X_s \gamma) \times 10^4$ & 3.2279          &  3.4300          &  0.1234          &  1.6374          \\
$\mathcal{B}(B_s \rightarrow \mu^+ \mu^-) \times 10^9$ & 3.1347          &  3.2000          &  0.0400          &  1.6308          \\
\hline
\multicolumn{3}{|l}{Total $\chi^2$}  & {\bf 0.76734}&  \\\hline
\end{tabular}
\end{center}
\end{table}

\begin{table}[h!]
\begin{center}
\begin{footnotesize}
 \caption[8]{ \mbox{{\bf Benchmark model YUd}}\\ (1/$\alpha_G, \, M_G, \, \epsilon_3, \ \lambda$) = ($25.97,\, 2.89 \times 10^{16}$ GeV, $\, -1.44$ \%, $0.605$),\\
($m_{16}, \, M_{1/2}, \, A_0, \, \mu(M_Z)$) = ($20000,\, 300, \,
-41171, \, 878$) GeV,\\
($(m_{H_d}/m_{16})^2, \, (m_{H_u}/m_{16})^2, \, \tan\beta$) = ($1.87,  \, 1.62, \, 49.67$)
\label{t:YUd} % end of caption
}
\end{footnotesize}
\begin{tabular}{|l|c|c|c|c|}
\hline
Observable & Fit value & Experimental Value & Pull & Error \\
\hline
$M_Z$ &              91.1876         &  91.1876         &  0.0000          &  0.4559          \\
$M_W$ &              80.5432         &  80.4360         &  0.2664          &  0.4025          \\
$1/\alpha_{em}$ &    137.1270        &  137.0360        &  0.1327          &  0.0000          \\
$G_{\mu} \times 10^5$ & 1.1704          &  1.1664          &  0.3426          &  0.0117          \\
$\alpha_3$ &         0.1183          &  0.1184          &  0.1036          &  0.0009          \\
\hline
$M_t$ &              173.7543        &  173.5000        &  0.1921          &  1.3238          \\
$m_b(m_b)$ &         4.1837          &  4.1800          &  0.1014          &  0.0366          \\
$m_{\tau}$ &         1.7762          &  1.7768          &  0.0682          &  0.0089          \\
\hline
$M_h$ &              122.41          &  125.30          &  0.9412          &  3.0676          \\
\hline
$\mathcal{B}(B \rightarrow X_s \gamma) \times 10^4$ & 3.2293          &  3.4300          &  0.1226          &  1.6374          \\
$\mathcal{B}(B_s \rightarrow \mu^+ \mu^-) \times 10^9$ & 3.1325          &  3.2000          &  0.0414          &  1.6308          \\
\hline
\multicolumn{3}{|l}{Total $\chi^2$}  & {\bf  1.171}&  \\\hline
\end{tabular}
\end{center}
\end{table}
\clearpage
\begin{table}[ht!]
\begin{center}
\begin{footnotesize}
 \caption[8]{ \mbox{{\bf Benchmark model YUe}}\\ (1/$\alpha_G, \, M_G, \, \epsilon_3, \ \lambda$) = ($25.99,\, 2.80 \times 10^{16}$ GeV, $\, -1.57$ \%, $0.603$),\\
($m_{16}, \, M_{1/2}, \, A_0, \, \mu(M_Z)$) = ($20000,\, 400, \,
-41088, \, 975$) GeV,\\
($(m_{H_d}/m_{16})^2, \, (m_{H_u}/m_{16})^2, \, \tan\beta$) = ($1.88,  \, 1.62, \, 49.62$)
\label{t:YUe} % end of caption
}
\end{footnotesize}
\begin{tabular}{|l|c|c|c|c|}
\hline
$M_Z$ &              91.1876         &  91.1876         &  0.0000          &  0.4559          \\
$M_W$ &              80.5546         &  80.4360         &  0.2946          &  0.4025          \\
$1/\alpha_{em}$ &    137.0434        &  137.0360        &  0.0109          &  0.0000          \\
$G_{\mu} \times 10^5$ & 1.1718          &  1.1664          &  0.4629          &  0.0117          \\
$\alpha_3$ &         0.1182          &  0.1184          &  0.1644          &  0.0009          \\
\hline
$M_t$ &              173.8117        &  173.5000        &  0.2355          &  1.3238          \\
$m_b(m_b)$ &         4.1840          &  4.1800          &  0.1098          &  0.0366          \\
$m_{\tau}$ &         1.7761          &  1.7768          &  0.0797          &  0.0089          \\
\hline
$M_h$ &              121.24          &  125.30          &  1.3219          &  3.0676          \\
\hline
$\mathcal{B}(B \rightarrow X_s \gamma) \times 10^4$ & 3.2057          &  3.4300          &  0.1370          &  1.6374          \\
$\mathcal{B}(B_s \rightarrow \mu^+ \mu^-) \times 10^9$ & 3.6184          &  3.2000          &  0.2566          &  1.6308          \\
\hline
\multicolumn{3}{|l}{Total $\chi^2$}  &  {\bf 2.23413}&  \\\hline
\end{tabular}
\end{center}
\end{table}

\begin{table}[ht!]
\begin{center}
\begin{footnotesize}
 \caption[8]{ \mbox{{\bf Benchmark model YUf}}\\ (1/$\alpha_G, \, M_G, \, \epsilon_3, \ \lambda$) = ($26,\, 2.5 \times 10^{16}$ GeV, $\, -1.16$ \%, $0.597$),\\
($m_{16}, \, M_{1/2}, \, A_0, \, \mu(M_Z)$) = ($20000,\, 600, \,
-41221, \, 664$) GeV,\\
($(m_{H_d}/m_{16})^2, \, (m_{H_u}/m_{16})^2, \, \tan\beta$) = ($1.87,  \, 1.62, \, 49.24$)
\label{t:YUf} % end of caption
}
\end{footnotesize}
\begin{tabular}{|l|c|c|c|c|}
\hline
$M_Z$ &              91.1876         &  91.1876         &  0.0000          &  0.4559          \\
$M_W$ &              80.5577         &  80.4360         &  0.3025          &  0.4025          \\
$1/\alpha_{em}$ &    136.8484        &  137.0360        &  0.2742          &  0.0000          \\
$G_{\mu} \times 10^5$ & 1.1729          &  1.1664          &  0.5616          &  0.0117          \\
$\alpha_3$ &         0.1183          &  0.1184          &  0.1345          &  0.0009          \\
\hline
$M_t$ &              173.6236        &  173.5000        &  0.0934          &  1.3238          \\
$m_b(m_b)$ &         4.1870          &  4.1800          &  0.1904          &  0.0366          \\
$m_{\tau}$ &         1.7757          &  1.7768          &  0.1234          &  0.0089          \\
\hline
$M_h$ &              119.65          &  125.30          &  1.8411          &  3.0676          \\
\hline
$BR(B \rightarrow X_s \gamma) \times 10^4$ & 3.1850          &  3.4300          &  0.1496          &  1.6374          \\
$BR(B_s \rightarrow \mu^+ \mu^-) \times 10^9$ & 3.1953          &  3.2000          &  0.0029          &  1.6308          \\
\hline
\multicolumn{3}{|l}{Total $\chi^2$}  &  {\bf 3.97248} &  \\
\hline
\end{tabular}
\end{center}
\end{table}

%%%%%%%%%%%%%%%%%%% BIBLIOGRAPHY %%%%%%%%%%%%%%%%%%%
%\clearpage
%\newpage

\bibliography{bibliography}

\bibliographystyle{utphys}

\end{document}